\documentclass[journal, 12pt, onecolumn, draftclsnofoot]{IEEEtran}
\pdfoutput=1
\IEEEoverridecommandlockouts

\usepackage{acronym}
\usepackage{amsmath,amssymb}
\usepackage[english]{babel}
\usepackage{bbm}
\usepackage{caption}
\usepackage{cite}
\usepackage[shortlabels,inline]{enumitem}
\usepackage{etoolbox}
\usepackage[T1]{fontenc}
\usepackage{graphicx}
\usepackage{hyperref}
\usepackage{ifthen}
\usepackage{cleveref}
\usepackage[utf8]{inputenc}
\usepackage{lipsum}
\usepackage{siunitx}
\usepackage{xcolor}

\def \arXiv {}

\newtheorem{proposition}{Proposition}
\newtheorem{corollary}{Corollary}
\newtheorem{remark}{Remark}

\newcommand{\kr}{\ensuremath{\kappa_r}}
\newcommand{\kad}{\ensuremath{\kappa_{a_0}}}
\newcommand{\ka}{\ensuremath{\kappa_a}}
\newcommand{\kd}{\ensuremath{\kappa_d}}
\newcommand{\xmin}{\ensuremath{x_{\mathrm{min}}}}
\newcommand{\xmax}{\ensuremath{x_{\mathrm{max}}}}
\newcommand{\ymin}{\ensuremath{y_{\mathrm{min}}}}
\newcommand{\ymax}{\ensuremath{y_{\mathrm{max}}}}
\newcommand{\zmin}{\ensuremath{z_{\mathrm{min}}}}
\newcommand{\zmax}{\ensuremath{z_{\mathrm{max}}}}

\newcommand{\anmp}{\alpha_n^{m,p}}
\newcommand{\anmpsq}{\ensuremath{\left(\anmp\right)^2}}
\newcommand{\lnmp}{\lambda_n^{m,p}}
\newcommand{\btm}{\beta_m}
\newcommand{\gp}{\gamma_p}
\newcommand{\hlx}{\kr}
\newcommand{\klx}{D}
\newcommand{\hrx}{\ka}
\newcommand{\krx}{D}
\newcommand{\hdx}{\kd}
\newcommand{\hly}{\ensuremath{\kappa_{1,y}}}
\newcommand{\kly}{k_{1,y}}
\newcommand{\hry}{\ensuremath{\kappa_{2,y}}}
\newcommand{\kry}{k_{2,y}}
\newcommand{\hlz}{\ensuremath{\kappa_{1,z}}}
\newcommand{\klz}{k_{1,z}}
\newcommand{\hrz}{\ensuremath{\kappa_{2,z}}}
\newcommand{\krz}{k_{2,z}}
\newcommand{\lx}{l_x}
\newcommand{\ly}{l_y}
\newcommand{\lz}{l_z}
\newcommand{\itot}{\ensuremath{i_{\textrm{tot}}}}

\newcommand{\dy}{\ensuremath{\mathrm{d}y}}
\newcommand{\dz}{\ensuremath{\mathrm{d}z}}

\newcommand{\ind}[2]{\ensuremath{\mathbbm{1}_{\lbrace #2 \rbrace}(#1)}}

\acrodef{DMC}{Diffusive Molecular Communication}
\acrodef{EH}{Energy Harvesting}
\acrodef{CIR}{channel impulse response}
\acrodef{ISI}{inter-symbol interference}
\acrodef{MC}{Molecular Communication}
\acrodef{BMI}{brain-machine interface}
\acrodef{GABA}{$\gamma$-aminobutyric acid}
\acrodef{NT}{neurotransmitter}
\acrodef{wrt}[w.r.t.]{with respect to}
\acrodef{wlog}[w.l.o.g.]{without loss of generality}
\acrodef{AMPA}{$\alpha$-amino-3-hydroxy-5-methyl-4-isoxazolepropionic acid}
\acrodef{ODE}{ordinary differential equation}

\makeatletter
\long\def\@makecaption#1#2{\ifx\@captype\@IEEEtablestring%
    \footnotesize\begin{center}{\normalfont\footnotesize #1}\\
        {\normalfont\footnotesize\scshape #2}\end{center}%
    \@IEEEtablecaptionsepspace
    \else
    \@IEEEfigurecaptionsepspace
    \setbox\@tempboxa\hbox{\normalfont\footnotesize {#1.}~~ #2}%
    \ifdim \wd\@tempboxa >\hsize%
    \setbox\@tempboxa\hbox{\normalfont\footnotesize {#1.}~~ }%
    \parbox[t]{\hsize}{\normalfont\footnotesize \noindent\unhbox\@tempboxa#2}%
    \else
    \hbox to\hsize{\normalfont\footnotesize\hfil\box\@tempboxa\hfil}\fi\fi}
\makeatother

\begin{document}
\bstctlcite{disable_url}    
\title{Synaptic Channel Modeling for DMC:\\ Neurotransmitter Uptake and Spillover in the Tripartite Synapse}
\author{\IEEEauthorblockN{Sebastian Lotter, Arman Ahmadzadeh, and Robert Schober}
    \thanks{This paper has been presented in part at the IEEE International Conference on Communications (ICC), 2020.
        This work was supported in part by the German Research Foundation under Project SCHO 831/9-1}}

\maketitle
\nocite{lotter20a}
\begin{abstract}
In \ac{DMC}, information is transmitted by diffusing molecules.
Synaptic signaling, as a natural implementation of this paradigm, encompasses functional components that, once understood, can facilitate the development of synthetic \ac{DMC} systems.
To unleash this potential, however, a thorough understanding of the synaptic communication channel based on biophysical principles is needed.
Since synaptic transmission critically depends also on non-neural cells, such understanding requires the consideration of the so-called tripartite synapse.
In this paper, we develop a comprehensive channel model of the tripartite synapse encompassing a three-dimensional, finite-size spatial model of the synaptic cleft, molecule uptake at the presynaptic neuron and at glial cells, reversible binding to individual receptors at the postsynaptic neuron, and spillover to the extrasynaptic space. 
Based on this model, we derive analytical time domain expressions for the \ac{CIR} of the synaptic \ac{DMC} system and for the number of molecules taken up at the presynaptic neuron and at glial cells, respectively.
These expressions provide insight into the impact of macroscopic physical channel parameters on the decay rate of the \ac{CIR} and the reuptake rate, and reveal fundamental limits for synaptic signal transmission induced by chemical reaction kinetics and the channel geometry.
Adapted to realistic parameters, our model produces plausible results when compared to previous experimental and simulation studies and we provide results from particle-based computer simulations to further validate the analytical model.
The proposed comprehensive channel model admits a wide range of synaptic configurations making it suitable for the investigation of many practically relevant questions, such as the impact of glial cell uptake and spillover on signal transmission in the tripartite synapse.    
\end{abstract}

\section{Introduction}
\acresetall
In nature, information exchange between biological entities, such as cells or organs, is often based on the release, propagation, and sensing of molecules.
This process is called \ac{MC}.
Although traditionally studied by biologists and medical scientists, it has recently also attracted interest from the communications research community \cite{nakano13}.
\ac{MC} is envisioned to open several exciting new application areas for which communication at cell level, either between synthetic cells \cite{freitas99} or between synthetic and natural cells \cite{veletic2019}, is required.

\ac{MC} systems in which molecules propagate solely via Brownian motion, referred to as \acf{DMC} systems, constitute a promising candidate for synthetic \ac{MC} implementations, as they require neither special infrastructure, nor external energy supply.
The design of synthetic \ac{DMC} systems, however, poses several challenges.

Firstly, in environments for which synthetic \ac{DMC} is intended, such as the human body, the amount of molecules available for transmission is typically limited and hence the employed communication scheme needs to be extremely energy efficient.
The concept of \ac{EH} enables ultra-low-power applications in traditional communications and several methods to leverage it also for \ac{DMC} have been proposed recently \cite{deng17,guo18,musa20}.
The models considered in \cite{deng17,guo18}, however, are based on free-space propagation and can thus not exploit the channel characteristics, while in \cite{musa20} a non-biological \ac{EH} mechanism is considered which may not be applicable in all \ac{DMC} environments.

Secondly, because Brownian motion is an undirected propagation mechanism, \ac{DMC} channels are typically dispersive and thus susceptible to \ac{ISI}.
This problem is well-known and, accordingly, several approaches for \ac{ISI} mitigation in \ac{DMC} have been proposed in the literature, including ISI-aware modulation schemes \cite{arjmandi17}, forward error-correction codes \cite{leeson12,shih13}, and equalization techniques \cite{tepekule15}.
Whereas \cite{arjmandi17,leeson12,shih13,tepekule15} focus mainly on the transmitter and the receiver, in \cite{noel14,heren15}, enzymatic degradation of information molecules in the channel is considered for the mitigation of \ac{ISI}.
The approach in \cite{noel14} was inspired by the neuromuscular junction.
While enzymatic degradation at the neuromuscular junction allows for later recovery of the degraded molecules, i.e., \ac{EH}, it acts non-selectively on all emitted molecules, hence also weakening the desired part of the molecular signal \cite{noel14}.
In this paper, we consider molecular information transmission between neurons in the presence of surrounding neuroglia.
In this natural DMC system, \ac{EH} and \ac{ISI} mitigation are achieved by presynaptic and glial cell uptake of \acp{NT} from the {\em synaptic cleft}, the narrow space between two communicating neurons.
The results derived in this paper suggest that glial cell uptake may provide a more selective channel clearance mechanism than enzymatic degradation, retaining the peak value but suppressing the undesired tail of the signal, cf.~Section~\ref{sec:results:glial_cell_uptake}.

It is known that the uptake of \acp{NT} is critical for synaptic communication; several severe neurological diseases, including amyotrophic lateral sclerosis (ALS), Alzheimer disease (AD), and ischemia are associated with {\em overactivation} of excitatory receptors due to dysfunctional reuptake \cite{danbolt01}. Furthermore, reuptake is also critical for the {\em supply with \acp{NT}} at the presynaptic neuron because neurons themselves cannot synthesize \acp{NT} \cite{bak06}.
Hence, through the lens of a communications engineer, \ac{ISI} mitigation and \ac{EH} are vital for natural synaptic transmission, and thus, the study of \ac{NT} reuptake is important for synthetic \ac{MC} system design.

From a design perspective, in turn, it is most relevant to understand how the synaptic channel parameters impact reuptake and under which conditions reliable and energy-efficient communication is possible.
Accordingly, in this work, we study \ac{EH} and \ac{ISI} mitigation in the tripartite synapse incorporating molecule uptake at neuroglia and at the presynaptic neuron, and the escape of \acp{NT} from the synaptic cleft ({\em spillover}).

Synaptic communication has been previously studied in the \ac{MC} literature for bipartite synapses in \cite{balevi13, malak13, liu14, veletic14, maham15, veletic16, veletic16b, khan2017, ramezani17b, veletic18, ramezani18, ramezani18a, ramezani18c, khan18, khan19, veletic20}, for tripartite synapses in \cite{mesiti15, mesiti15a, veletic16, veletic20}, and for an artificial synapse in \cite{bilgin17}.
In most of these works, the synaptic cleft is modeled as ``black box'', and, accordingly, \ac{NT} propagation is modeled via a Bernoulli distributed random variable \cite{veletic16, veletic18, veletic20} or phenomenological models for the postsynaptic response based on the so-called alpha-function \cite{malak13, liu14, ramezani17b, ramezani18, khan18, khan19} or more general ``functional relationships'' \cite{mesiti15} are adopted.
In \cite{balevi13, veletic15, veletic16b}, diffusive \ac{NT} propagation is modeled as free-space propagation.
In \cite{khan2017}, the synaptic cleft in the bipartite synapse is modeled as three-dimensional region bounded by two parallel infinite planes.
However, experimental evidence suggests that by neglecting diffusion barriers in the synaptic cleft as in \cite{khan2017} the synaptic \ac{NT} concentration is likely to be underestimated as compared to experimental data \cite{clements96}.
Furthermore, in \cite{bilgin17}, a closed three-dimensional cuboid was proposed as a geometrical model for an artificial synapse.
The model proposed in \cite{bilgin17} allows for presynaptic reuptake of \acp{NT} and incorporates detailed kinetics for the main postsynaptic receptors.
However, \ac{NT} uptake at glial cells is not considered in \cite{bilgin17}.
In addition, the mathematical model proposed in \cite{bilgin17} is highly non-linear, rendering a deeper analysis impossible.
In \cite{liu14}, synaptic \ac{ISI} is studied using an alpha-function model for the molecular signaling in the synaptic cleft.
Spillover in the absence of glial cells has been studied in \cite{veletic18} modeling \ac{NT} propagation as Bernoulli distributed random variable.

In this work, we propose a comprehensive, three-dimensional analytical channel model for the tripartite synapse based on the diffusion equation and encompassing a finite-size spatial model of the synaptic cleft, molecule uptake at the presynaptic neuron and at glial cells, and reversible binding to individual receptors at the postsynaptic neuron.
While existing communication theoretic studies of the tripartite synapse \cite{mesiti15, mesiti15a, veletic16, veletic2019} focus on the astrocyte-neuron communication in terms of gliotransmission, i.e., astrocytic glutamate release in response to the activation of intracellular second messengers like $\mathrm{Ca}^{2+}$ and IP3 by synaptic glutamate, we adopt the view of the astrocyte as an insulator shielding the synapse from synaptic crosstalk and being responsible for clearing the \ac{MC} channel from residual \acp{NT} \cite{nedergaard12}.
In contrast to previous works on synaptic communication in the \ac{MC} literature, our model provides analytical insight into the spatio-temporal concentration profile of the \acp{NT} in the synaptic cleft in the presence of glial cell uptake and reversible receptor binding.

Based on the proposed model, combing tools from complex analysis and orthogonal function space decomposition \cite{ozisik13}, we first derive analytical time domain expressions for the \ac{CIR} of the synaptic \ac{DMC} system and for the number of molecules taken up at the presynaptic neuron and at glial cells, respectively.
Then, we use these expressions to characterize the fundamental limits that the channel geometry and the chemical reaction kinetics impose on the decay rate of the \ac{CIR} and the reuptake rate, respectively.
Our analysis shows that there exists an upper bound on the \ac{CIR} decay resulting from the channel geometry which is independent of the reaction kinetics and, similarly, it shows the existence of a second upper bound induced by the reaction kinetics which is independent of the channel geometry.
Furthermore, we also examine the impact of the individual macroscopic physical channel parameters, such as the presynaptic reuptake rate, on the asymptotic signal decay.
Thus, our model provides a framework for understanding the fundamental trade-offs between different synaptic channel parameters limiting signal transmission in the tripartite synapse in the presence of glial cell molecule uptake and spillover.
To the best of the authors' knowledge, the presented analytical model is the first such model in the \ac{MC} literature that simultaneously takes into account reversible receptor binding and presynaptic and glial \ac{NT} uptake for a finite-sized three-dimensional model of the tripartite synapse.
Furthermore, the combination of mathematical tools used to tackle the considered problem has not been employed previously in the \ac{MC} literature.

To ensure the biological significance of our model, realistic values for the various model parameters are adopted and the model outcome is compared with results from experimental studies.
Our analytical results are further validated by particle-based computer simulations and experimentally shown to provide an unbiased estimator for the \ac{CIR}.
When applied to different glial cell configurations, our model provides interesting novel insights into the potential of glial cell molecule uptake for \ac{ISI} mitigation.

The mathematical model presented in this paper is a generalization of the model previously reported in \cite{lotter20a} from one to three spatial dimensions.
Since the one-dimensional model in \cite{lotter20a} does not address the impact of molecule uptake by glial cells and spillover, respectively, the three-dimensional model presented in this paper is far more general and hence allows for richer analyses.

The remainder of this paper is organized as follows:
In Section~\ref{sec:model}, we state the system model and the main assumptions used in Section~\ref{sec:cir} to derive the time- and space-dependent synaptic molecule concentration, the \ac{CIR}, and the number of uptaken particles.
In Section \ref{sec:cir_decay_rate}, different regimes for the \ac{CIR} decay and the molecule reuptake rate are specified, and the asymptotic impact of the channel parameters is investigated.
The particle-based simulator design is outlined in Section~\ref{sec:pb_sim}, and numerical results are presented in Section~\ref{sec:results}.
Finally, the main findings are summarized in Section~\ref{sec:conclusion}.

\section{System Model}
\label{sec:model}
\subsection{Biological Background}\label{sec:model:biological_background}
In synaptic neurotransmission, the presynaptic neuron (transmitter) encodes a sequence of electrical impulses (data stream) into a spatio-temporal \ac{NT} release pattern (molecular signal) which propagates through the synaptic cleft (channel) and is finally received by the postsynaptic neuron (receiver) where the \acp{NT} activate membrane receptors.
In addition, trans-membrane transporter proteins at the presynaptic neuron and at surrounding neuroglia clear signaling molecules from the synaptic space, enabling recycling and future reuse \cite{bak06}.
In this way, the synaptic transmission is effectively terminated and, in addition, \acp{NT} are prevented from escaping from the synapse and causing interference at neighboring synapses (spillover) \cite{nielsen04,nedergaard12}.
Depending on the specific type of synapse, the reuptake at either the presynaptic neuron \cite{hasegawa06} or at neuroglia \cite{bak06} can be dominant.
In this work, we model synaptic communication as single-input single-output (SISO) system.
Hence, our model does not account for interference possibly caused by \acp{NT} released at neighboring synapses ({\em spill-in}).

\subsection{Geometry and Assumptions}\label{sec:model:geometry_and_assumptions}
\begin{figure*}
    \centering
    \includegraphics[width=.65\textwidth]{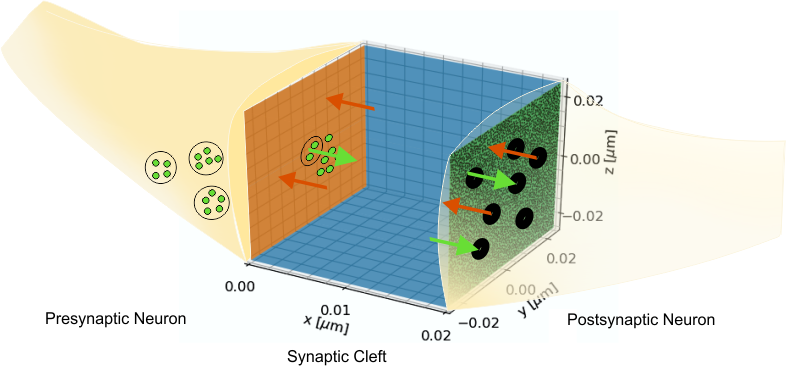}\vspace*{-3 mm}
    \caption{Model synapse. \acp{NT} enclosed in vesicles at the presynaptic neuron are released into the synaptic cleft, propagate by Brownian motion, and activate receptors at the postsynaptic neuron. Binding to postsynaptic receptors is reversible. Furthermore, particles are reuptaken (and recycled) at the presynaptic neuron. The cuboid represents an abstraction of the synaptic cleft. Its orange and green surfaces represent the membrane of the pre- and postsynaptic neuron, respectively. The blue faces and the two faces opposite to the blue ones which are transparent in this figure, may represent the cell membrane of astroglia encapsulating the synapse, but are in principle configurable to represent any kind of reflective, partially absorbing, or fully absorbing boundary.}\vspace*{-4mm}
    \label{fig:channel}
\end{figure*}

The synaptic cleft is abstracted in our model as a three-dimensional rectangular cuboid (see Fig.~\ref{fig:channel}).
This simplification appears biologically meaningful when compared to images of hippocampal synapses obtained by electron microscopy \cite[Fig.~2(H)]{pickel14_1}.
Moreover, it is analytically more tractable than the actual non-regular shaped synaptic domain, while still retaining its characteristic features.
A similar geometric model is used in \cite{khan2017}, with the difference that the model considered here is bounded in all dimensions, while in \cite{khan2017} only one dimension is bounded.

Formally, we characterize the domain of the synaptic cleft in Cartesian coordinates as follows
\begin{equation}
\Omega = \{(x,y,z) \vert \xmin \leq x \leq \xmax, \ymin \leq y \leq \ymax, \zmin \leq z \leq \zmax \}, \label{eq:domain}
\end{equation}
and the concentration of molecules in $\si{\per\cubic\micro\meter}$ at any time $t$ at any location within the box defined by \eqref{eq:domain} as $C(x,y,z,t)$.
To ease notation, we follow the convention to suppress the dependence of $C(x,y,z,t)$ on $x$, $y$, $z$, and $t$ in the notation whenever possible, i.e., we simply write $C$ instead.

Then, the presynaptic and postsynaptic membranes are located at $x=\xmin$ and $x=\xmax$, respectively.
The other system boundaries, $y=\ymin$, $y=\ymax$, $z=\zmin$, and $z=\zmax$, can be configured depending on the context the model is used in.
Here, as we are interested in reuptake in the tripartite synapse, they will mostly represent the membranes of the glial cell surrounding the synapse.
A different use case, also covered by this model, is the evaluation of the number of \acp{NT} that (irreversibly) leave the synapse and possibly cause interference at other synapses, i.e., spillover \cite{nielsen04}.

To derive the (dimensionless) impulse response of the synaptic channel, $h(t)$, we consider instantaneous release of $N$ particles at time $t=t_0$ at location $(x,y,z)=(x_0,y_0,z_0)\in \Omega$.
$h(t)$ is then given as the number of particles adsorbed to the receptors of the postsynaptic membrane as a function of time.
For our analysis, we make four assumptions:
\begin{enumerate}[{A}1)]
    \item All adsorption and desorption coefficients as well as the diffusion coefficient are constant.\label{ass:1}
    \item Particles diffuse independently of each other and the competition of receptors and transporters for ligands is negligible.\label{ass:2}
    \item The receptors at the postsynaptic membrane are uniformly distributed over the plane $x=\xmax$.\label{ass:3}
    \item Reversible adsorption to individual receptors with intrinsic binding rate $\kad$ in $\si{\micro\meter\per\micro\second}$ and intrinsic unbinding rate $\kd$ in $\si{\per\micro\second}$ can be treated equivalently as reversible adsorption to a homogeneous surface with effective binding rate $\ka$ in $\si{\micro\meter\per\micro\second}$ and the same unbinding rate, $\kd$.\label{ass:4}
\end{enumerate}
While the adsorption, desorption, and uptake rates can in reality change over time, in particular due to the effects of synaptic plasticity \cite{zucker14, danbolt01}, \ref{ass:1} is a reasonable assumption if the time frame under consideration is sufficiently short.
\ref{ass:2} can be justified for low molecule concentrations \cite{noel15,cussler09} and small-sized receptors \cite{ahmadzadeh16}. Together with \ref{ass:1} it renders the system under consideration linear and time-invariant, such that the derivation of the \ac{CIR} is actually meaningful.
By assuming that \acp{NT} diffuse and bind independent of each other, we neglect, in particular, saturation effects at the postsynaptic receptors and at the \ac{NT} transporters resulting from the competition of a finite number of receptors/transporters for molecules.
In the case of postsynaptic receptors, this assumption is justified under physiological conditions by experimental observations \cite{mcallister00}.
For \ac{NT} transporters, we assume that the uptake is fast relative to the binding of molecules to transporters and, hence, the impact of saturation effects on molecule uptake is negligible.
In reality, synaptic transmission can be non-linear and time-variant due to the effects of saturation and synaptic plasticity.
While this is a challenging topic for future research, we do not consider it here in order to keep our model analytically tractable.
\ref{ass:3} is reasonable if we assume that the so-called postsynaptic density \cite{pickel14_1}, the part of the postsynaptic membrane which contains most receptors, extends over the entire postsynaptic surface under consideration.
Finally, \ref{ass:4} is not obvious.
For irreversible adsorption to an otherwise reflective surface covered by partially adsorbing disks, boundary homogenization, i.e., the assumed equivalence, has been justified in \cite{zwanzig1991} and quantitatively refined with results from computer simulation in \cite{berezhkovskii04}.
However, it is not clear if similar techniques can be applied for reversible reactions, too.
There are two main issues to consider here.
First, the steady-state fluxes are substantially different;
the net flux at a reversibly adsorbing boundary is zero at steady-state, while for irreversible adsorption it is nonzero.
Second, desorption alters the spatial concentration profile of particles near the boundary; particles are more concentrated near receptors compared to irreversible adsorption \cite{szabo91}.
The first issue can be resolved by calibrating the effective adsorption coefficient to the homogenized surface such that it correctly reproduces the steady-state flux to the patchy surface (see Section~\ref{sec:pb_sim}), while the second issue can be resolved in a biologically plausible manner by assuming that particles may not re-adsorb immediately after unbinding \cite{zarnitsyna07}.

According to Fick's second law of diffusion, (average) Brownian particle motion can be described by the following partial differential equation:
\begin{equation}
\frac{\partial{}C}{\partial{}t} = D \Delta C, \label{eq:diff_eq}
\end{equation}
where $D > 0$ denotes the particle diffusion coefficient in $\si{\micro\meter\squared\per\micro\second}$, and $\Delta$ is the Laplace operator in Cartesian coordinates, defined as $\Delta f = \frac{\partial{}^2 f}{\partial{}x^2} + \frac{\partial{}^2 f}{\partial{}y^2} + \frac{\partial{}^2 f}{\partial{}z^2}$.

\subsection{Pre- and Postsynaptic Neurons}\label{sec:model:pre_and_postsynaptic_neurons}
Particle reuptake at the presynaptic neuron is modeled as irreversible adsorption to the homogeneous boundary at $x=\xmin$ (possibly after appropriate boundary homogenization) with reuptake coefficient $\kr$ in $\si{\micro\meter\per\micro\second}$.
To simplify notation, we set $\xmin=0$ and $\xmax=\lx$.
Then, the boundary condition modeling presynaptic reuptake is given as \cite{collins49}
\begin{equation}
D\frac{\partial C}{\partial x} =  \kr C, \text{ at } x=0,\label{eq:pab_lb}
\end{equation}
where $\kr \geq 0$.
A boundary condition of the form in \eqref{eq:pab_lb} is referred to as {\em boundary condition of the third kind} \cite{ozisik13}.

To model reversible adsorption to the postsynaptic boundary, we adopt the {\em backreaction} boundary condition \cite{agmon84} and obtain
\begin{equation}
D\frac{\partial C}{\partial x} = - \kappa_a C - \kappa_d \int_{0}^{t} D\frac{\partial C}{\partial x} \mathrm{d}\tau, \text{ at } x = \lx, \label{eq:pab_rb}
\end{equation}
where $\ka > 0$ and $\kd \geq 0$.
Although $\ka = 0$ would technically be also valid, the signal at the postsynapse would in this case be identical to $0$, so we restrict our considerations to $\ka > 0$.

By modeling postsynaptic binding using \eqref{eq:pab_rb} and assuming \ref{ass:2} in Section~\ref{sec:model:geometry_and_assumptions}, we adopt a simplistic two-state (bound/unbound) model for the postsynaptic receptors that is, however, accurate for the main ionotropic receptors at excitatory glutamatergic synapses \cite{destexhe94}.
As metabotropic postsynaptic receptors mainly mediate synaptic adaptation processes, they are not considered in this work.
For a more realistic (non-linear) receptor model, we refer the reader to \cite{bilgin17}.

\subsection{Molecule Uptake at Glial Cell}

For the boundaries at $y=\ymin$, $y=\ymax$, $z=\zmin$, and $z=\zmax$, we use boundary conditions of the third kind, because they provide most flexibility to our model.
In particular, perfectly absorbing and perfectly reflecting surfaces are covered as special cases.
For \ac{NT} uptake by glial cells, the situation is equivalent to the presynaptic membrane and thus this choice is justified by the same argument as in Section \ref{sec:model:pre_and_postsynaptic_neurons}.

Although we focus in this paper on the modeling of the tripartite synapse, the proposed model can as well be applied to bipartite synapses if either particles that leave the synaptic cleft are unlikely to come back or presynaptic reuptake provides the dominant clearance mechanism for \acp{NT}. In the first case, escape of \acp{NT} from the synaptic cleft can be modeled as absorption in $y$ and $z$ \cite{trommershauser99}.
The second case is most relevant for synapses with diffusion barriers in $y$ and $z$ which keep the \acp{NT} within reach for presynaptic reuptake, e.g.~the Ribbon synapse \cite{hasegawa06}.
In this case, the diffusion barriers in $y$ and $z$ are modeled with reflective boundaries and our model collapses in terms of the \ac{CIR} to the one-dimensional model proposed in \cite{lotter20a}.

Setting $\ymin=0$, $\ymax=\ly$, $\zmin=0$, and $\zmax=\lz$, this yields
\begin{align}
k_{1,j} \frac{\partial C}{\partial j} &= \kappa_{1,j} C \textrm{ at } j=0, j \in \{y,z\},\label{eq:rb_l} \\
k_{2,j} \frac{\partial C}{\partial j} &= -\kappa_{2,j} C \textrm{ at } j=l_j, j \in \{y,z\},\label{eq:rb_r}
\end{align}
where $k_{i,j} \in \{0,D\}$, $\kappa_{i,j} \geq 0$ in $\si{\micro\meter\per\micro\second}$ is the adsorption coefficient (possibly after boundary homogenization) at boundary $j=0$ for $i=1$ and $j=l_j$ for $i=2$, and $k_{i,j}$ and $\kappa_{i,j}$ are not both to vanish at the same time, i.e., $k_{i,j} + \kappa_{i,j} > 0$.

\subsection{Particle Release}\label{sec:model:part_rel}
We consider the instantaneous release of $N$ particles at $t_0$.
Without loss of generality (w.l.o.g.)\acused{wlog}, we set $t_0=0$ and $N=1$.
The former assumption is not restrictive because the system is time-invariant.
The latter one is not restrictive because the system is linear and, hence, results can simply be rescaled by $N$.

The instantaneous release of $1$ particle is then modeled as the initial value
\begin{equation}
C_0 = C(x,y,z,0) = \delta(x-x_0)\delta(y-y_0)\delta(z-z_0), (x_0,y_0,z_0) \in \Omega, \label{eq:pab_ic}
\end{equation}
where $\delta(\cdot)$ denotes the Dirac delta function.

In reality, when we consider multiple vesicular releases of \acp{NT} into the synaptic cleft, these do not always happen at the same release position, neither is the release instantaneous.
However, as we will derive the {\em Green's function} of the system, our result can be easily adapted to more realistic release models \cite{clements96} by using the theory of Green's functions \cite{carslaw86}.

We note that, in this paper, we focus on the modeling of the synaptic channel in terms of the effects relevant for the \ac{NT} concentration in the synaptic cleft {\em after the fusion of the vesicle with the presynaptic membrane}.
For an end-to-end model of synaptic communication, a more detailed model of the transmitter, incorporating e.g.~the dependence of vesicular release on the availability of \acp{NT}, would be required.

The most important model parameters used in the remainder of this paper are summarized for further reference in Table~\ref{tab:sim_params}.

\begin{table}
    \centering
    \caption{Model parameters and default values for particle-based simulation.}
    \footnotesize
    \begin{tabular}{| l | c | r | l |}
        \hline Parameter & Description & Default Value & Reference(s)\\ \hline
        $D$ & Diffusion coefficient & $\SI{3.3e-4}{\micro\meter\squared\per\micro\second}$ & \cite{nielsen04} \\ \hline
        $\kly,\kry,\klz,\krz$ & Boundary flux coefficient in $y$ and $z$ & $\SI{3.3e-4}{\micro\meter\squared\per\micro\second}$ & \cite{nielsen04} \\ \hline
        $N$ & Number of released \acp{NT} & $\SI{3000}{}$ & \cite{savtchenko13}\\ \hline
        $\lx$ & Cleft width in $x$ & $\SI{0.02}{\micro\meter}$ & \cite{ventura99}\\ \hline
        $\ly, \lz$ & Cleft widths in $y$ and $z$ & $\SI{0.15}{\micro\meter}$ & \cite{ventura99}\\ \hline
        $\hlx$ & Presynaptic reuptake rate & $1.3 \times 10^{-6}$~$\si{\micro\meter\per\micro\second}$ & \cite{holmes95}\\ \hline
        $\hrx$ & Postsynaptic adsorption rate & $1.5 \times 10^{-5}$~$\si{\micro\meter\per\micro\second}$ & \cite{holmes95,jonas93}\\ \hline
        $\hdx$ & Postsynaptic desorption rate & $8.5 \times 10^{-3}$~$\si{\per\micro\second}$ & \cite{holmes95,jonas93}\\ \hline
        $\hly,\hry,\hlz,\hrz$ & Adsorption coefficients at glial cell & $2.6 \times 10^{-5}$~$\si{\micro\meter\per\micro\second}$ & \cite{wadiche98, lehre98}\\ \hline
        $\Delta t$ & Simulation time step & $\SI{0.01}{\micro\second}$ & \\ \hline
        $r$ & Postsynaptic receptor radius & $\SI{0.7}{\nano\meter}$ & \\ \hline
        $\rho$ & Postsynaptic receptor coverage & $\SI{0.15}{}$ & \\ \hline
        $|\alpha^{1,1}_1|$, $\beta_1$, $\gamma_1$ & Dominant eigenvalues in $x$, $y$, and $z$ & $\SI{0.08}{\per\micro\meter}$, $\SI{1.02}{\per\micro\meter}$, $\SI{1.02}{\per\micro\meter}$ & \\ \hline
    \end{tabular}\vspace*{-4mm}
    \label{tab:sim_params}
\end{table}

\subsection{Channel Impulse Response and Number of Reuptaken Particles}

The number of particles adsorbed to the postsynaptic boundary at time $t$ equals the sum of all positive and negative fluxes over this boundary in the interval $\lbrack0;t\rbrack$.
Thus, once the solution to \labelcref{eq:diff_eq,eq:pab_lb,eq:pab_rb,eq:rb_l,eq:rb_r,eq:pab_ic} is found, the \ac{CIR}, $h(t)$, can be obtained as the accumulated net flux over the entire postsynaptic boundary,
\begin{equation}
h(t) = \int\limits_{0}^{t}\int\limits_{0}^{\ly}\int\limits_{0}^{\lz} -D\left.\frac{\partial C(x,y,z,\tau)}{\partial x}\right|_{x=\lx} \mathrm{d}z\mathrm{d}y\mathrm{d}\tau. \label{eq:def_h}
\end{equation}

For our further analysis, we denote $h(t)$ in the special case of reflective boundaries in $y$ and $z$, i.e., $\kappa_{i,j} = 0, \forall i \in \{1,2\}, j \in \{y,z\}$, by $h^r(t)$.

Similarly, the number of reuptaken particles, $i(t)$, is obtained as the cumulative flux over the presynaptic boundary,
\begin{equation}
i(t) = \int\limits_{0}^{t}\int\limits_{0}^{\ly}\int\limits_{0}^{\lz} D\left.\frac{\partial C(x,y,z,\tau)}{\partial x}\right|_{x=0} \mathrm{d}z\mathrm{d}y\mathrm{d}\tau.\label{eq:def_i}
\end{equation}

The number of reuptaken particles is a relevant quantity, because, first, it provides an upper bound on the \ac{ISI} caused by the signaling molecules remaining in the synaptic cleft after transmission and, second, since neurons cannot synthesize \acp{NT} on their own \cite{bak06}, it provides a lower bound on the time that the presynapse needs to recover from signal transmission, such that it can refill the pool of its readily releasable vesicles.

If the boundaries in $y$ and $z$ are reflective ($\kappa_{i,j} = 0, \forall i \in \{1,2\}, j \in \{y,z\}$) and adsorption to the postsynaptic boundary is actually reversible ($\kd > 0$), eventually all particles are reuptaken.
In this special case, we denote $i(t)$ as $i^r(t)$ and have
\begin{equation}
\lim\limits_{t \to \infty} i^r(t) = 1\label{eq:i_cdf}.
\end{equation}
As $D\left.\frac{\partial C}{\partial x}\right|_{x=0} = \kr C \geq 0$, we can then interpret $i^r(t)$ as cumulative distribution function for the probability that a particle released at $t_0=0$ is reuptaken in the interval $[0;t]$.

Furthermore, for the investigation of glial cell molecule uptake, we define the number of molecules uptaken at the glial cell as
\begin{align}
i_{-}(t) &= \int\limits_{0}^{t}\int\limits_{0}^{\lx}\int\limits_{0}^{\lz} \kly\left.\frac{\partial C(x,y,z,\tau)}{\partial y}\right|_{y=0} - \kry\left.\frac{\partial C(x,y,z,\tau)}{\partial y}\right|_{y=\ly} \mathrm{d}z\mathrm{d}x\mathrm{d}\tau\nonumber\\
&{}+\int\limits_{0}^{t}\int\limits_{0}^{\lx}\int\limits_{0}^{\ly} \klz\left.\frac{\partial C(x,y,z,\tau)}{\partial z}\right|_{z=0} - \krz\left.\frac{\partial C(x,y,z,\tau)}{\partial z}\right|_{z=\lz} \mathrm{d}y\mathrm{d}x\mathrm{d}\tau,\label{eq:def_ig}
\end{align}
and the total number of uptaken particles as
\begin{equation}
\itot(t) = i(t) + i_{-}(t).\label{eq:def_itot}
\end{equation}

If $\hdx > 0$, as particles cannot leave the system over any other boundary, we have $\lim\limits_{t \to \infty} \itot(t) = 1$ and $\itot(t)$ is the cumulative distribution function for the probability that a particle is uptaken in the interval $[0;t]$ at either the glial cell or the presynaptic neuron.

\section{Analytical Channel Model}
\label{sec:cir}
\subsection{Molecule Concentration}
If our system model would comprise only boundaries of the third kind, the solution of the three-dimensional problem would be just a multiplication of solutions to the corresponding one-dimensional problems and could easily be obtained from the literature \cite{carslaw86}.
However, the particular challenge in solving \labelcref{eq:diff_eq,eq:pab_lb,eq:pab_rb,eq:rb_l,eq:rb_r,eq:pab_ic} is the temporal coupling induced by the {\em backreaction} boundary condition \eqref{eq:pab_rb}.

\begin{proposition}\label{prop:C}        
    Let $\hlx, \hdx \geq 0$, $\hrx > 0$, $k_{i,j} \in \{0,D\}$, $\kappa_{i,j} \geq 0$,  and $k_{i,j} + \kappa_{i,j} > 0, \forall i \in \{1,2\}, j \in \{y,z\}$.
    Then, the unique solution to \labelcref{eq:diff_eq,eq:pab_lb,eq:pab_rb,eq:rb_l,eq:rb_r,eq:pab_ic} is
    \begin{align}
    C(x,y,z,t) &= \sum_{m=1}^{\infty}\sum_{p=1}^{\infty} \left[\frac{\hdx}{\hrx + \lx \hdx} \frac{1}{\ly \lz} \ind{\kr}{0}\ind{\btm}{0}\ind{\gp}{0}\right.\nonumber\\ 
    &\left. + \sum_{n=1}^{\infty} \mathcal{A}_n^{m,p} \mathcal{B}_m \mathcal{C}_p C_0(x_0,y_0,z_0) C_{n,x}^{m,p}(x) C_{m,y}(y) C_{p,z}(z) \exp(-\lnmp t)\right],\label{eq:C}
    \end{align}
    where
    \begin{align}
    C_{n,x}^{m,p}(x) &= \hlx \sin(\anmp x) + \klx \anmp \cos(\anmp x),\label{eq:def_Cmpnx}\\
    C_{m,y}(y) &= \hly \sin(\btm y) + \kly \btm \cos(\btm y), \\
    C_{p,z}(z) &= \hlz \sin(\gp z) + \klz \gp \cos(\gp z),\\
    C_0(x_0,y_0,z_0) &= C_{n,x}^{m,p}(x_0)C_{m,y}(y_0)C_{p,z}(z_0),\label{eq:def_C0} \\
    \mathcal{A}_n^{m,p} &= 2\left[(\anmp)^2 \krx^2 (\lnmp - \hdx)^2 + (\lnmp)^2 \hrx^2\right] / \nonumber \\
    &\{(\lnmp - \hdx)\left[D^3 (\anmp)^2(\hlx (\lnmp - \hdx) + \hrx \lnmp) + \hlx^2 \hrx \krx \lnmp\right]\nonumber \\
    &+ 2 \hlx^2 \hrx \hdx D^2 (\anmp)^2 + \hlx \hrx^2 \klx (\lnmp)^2 + 2 \hrx \hdx D^4 (\anmp)^4\nonumber \\
    &+l_x \left[\hlx^2 \krx^2 (\anmp)^2 (\lnmp - \hdx)^2 + \hrx^2 \klx^2 (\anmp)^2 (\lnmp)^2\right.\nonumber \\
    &+ \left.D^4 (\anmp)^4 (\lnmp - \hdx)^2 + \hlx^2 \hrx^2 (\lnmp)^2\right]\},\label{eq:def_Anmp}\\
    \mathcal{B}_m &= (2-\ind{\btm}{0})(\btm^2 \kry^2 + \hry^2)/\left[l_y(\hly^2+\kly^2\btm^2)(\hry^2+\kry^2\btm^2)\right. \nonumber \\
    &+\left.\hly \kly (\hry^2+\kry^2\btm^2)+ \hry \kry (\hly^2+\kly^2\btm^2)\right],\label{eq:def_Btm} \\
    \mathcal{C}_p &= (2-\ind{\gp}{0})(\gp^2 \krz^2 + \hrz^2)/\left[l_z(\hlz^2+\klz^2\gp^2)(\hrz^2+\krz^2\gp^2)\right. \nonumber \\
    &+\left.\hlz \klz (\hrz^2+\krz^2\gp^2)+ \hrz \krz (\hlz^2+\klz^2\gp^2)\right],\label{eq:def_Gp} \\
    \lnmp &= D((\anmp)^2+\btm^2+\gp^2),\label{eq:def_lnmp}
    \end{align}
    $\anmp$ are the non-zero roots of
    \begin{equation}
    \tan(\alpha l_x) = \frac{\alpha D\left[D(\alpha^2+\btm^2+\gp^2)(\hlx+\hrx)-\hlx\hdx\right]}{D (\alpha^2+\btm^2+\gp^2)(\alpha^2 D^2-\hlx\hrx)-\alpha^2\hdx D^2},\label{eq:def_an}
    \end{equation}
    with either positive real part or positive imaginary part, $\btm$ and $\gp$ are the positive, or, if $\hly=\hry=0$ $(\hlz=\hrz=0)$, the real, non-negative, roots of
    \begin{equation}
    \tan(\beta l_y) = \frac{\beta(\hly\kry+\hry\kly)}{\beta^2\kly\kry-\hly\hry},\label{eq:def_btm}
    \end{equation}
    and
    \begin{equation}
    \tan(\gamma l_z) = \frac{\gamma(\hlz\krz+\hrz\klz)}{\gamma^2\klz\krz-\hlz\hrz},\label{eq:def_gp}
    \end{equation}
    respectively, and $\ind{x}{S}$ denotes the indicator function, $\ind{x}{S} = 1$ if $x \in S$, $\ind{x}{S} = 0$ otherwise.
\end{proposition}

\begin{IEEEproof}
\ifx \arXiv \undefined
    Due to space constraints, we can provide only a sketch of the proof.
    The full proof can be found in an extended version of this paper, see \cite{lotter20}.
    To derive the solution to \labelcref{eq:diff_eq,eq:pab_lb,eq:pab_rb,eq:rb_l,eq:rb_r,eq:pab_ic}, we exploit the spatial separability of the diffusion equation in the Cartesian coordinate system and rewrite $C(x,y,z,t)$ as linear combination of orthonormal eigenfunctions of the Laplace operator in $y$ and $z$ with coefficients being functions of $x$ and $t$.
    Then, we use the Laplace transform \ac{wrt}~$t$ as defined in the Appendix of \cite{lotter20a} to obtain a second order linear \ac{ODE} in $x$ for each of these coefficients.
    The solutions of these \acp{ODE} are transformed back into the time domain by application of the residue theorem and, finally, $C(x,y,z,t)$ as stated in Proposition \ref{prop:C} is obtained.
\else
    Please refer to the Appendix.
\fi
\end{IEEEproof}

\begin{remark}
The aforementioned coupling introduced by \eqref{eq:pab_rb} becomes apparent in the dependence of the $\anmp$ on $\btm$ and $\gp$.
In fact, if $\hdx=0$, i.e.,~\eqref{eq:pab_rb} collapses to an ordinary boundary of the third kind, \eqref{eq:def_Anmp} and \eqref{eq:def_an} become independent of $\btm$ and $\gp$, and $C(x,y,z,t)$ factorizes into a product of one-dimensional solutions.
\end{remark}

\begin{remark}
Although imaginary values are allowed for $\anmp$, $C(x,y,z,t)$ is always a real-valued function, because the trigonometric functions occurring in \eqref{eq:C} when evaluated at purely imaginary arguments equal their hyperbolic counterparts evaluated at real arguments and the leftover imaginary parts cancel each other out.
\end{remark}

\begin{remark}
$C(x,y,z,t)$ is called the {\em Green's function} of the boundary value problem \labelcref{eq:diff_eq,eq:pab_lb,eq:pab_rb,eq:rb_l,eq:rb_r} \cite{carslaw86}.
\end{remark}

We index the sets $\lbrace \btm \rbrace$ and $\lbrace \gp \rbrace$ ascendingly, such that $\beta_1$ and $\gamma_1$ are the smallest admissible solutions of \eqref{eq:def_btm} and \eqref{eq:def_gp}, respectively.
Furthermore, we remark that \eqref{eq:def_an} admits at most one imaginary solution with positive imaginary part and we index the real roots of \eqref{eq:def_an} ascendingly, starting with index $1$ if \eqref{eq:def_an} does {\em not} admit an imaginary root and starting with index $2$ otherwise.
In the latter case, we index the imaginary root of \eqref{eq:def_an} with index $1$.

\subsection{CIR and Number of Reuptaken Particles}
Next, we use Proposition \ref{prop:C} to derive the \ac{CIR} and the number of reuptaken particles.
\subsubsection{CIR}
First, we consider the \ac{CIR}.
\begin{corollary}
The \ac{CIR}, $h(t)$, as defined in \eqref{eq:def_h} can be obtained from \eqref{eq:C} as
\begin{align}
h(t) &= \sum_{n=1}^{\infty}\sum_{m=1}^{\infty}\sum_{p=1}^{\infty} \mathcal{A}_n^{m,p} \mathcal{B}_m \mathcal{C}_p C_0(x_0,y_0,z_0)(\hlx \anmp \cos(\anmp \lx) - \klx \anmpsq \sin(\anmp \lx)) \nonumber\\
&\times \left[\kly \sin(\btm \ly) + \frac{\hly}{\btm} (1 - \cos(\btm \ly))\right]\left[\klz \sin(\gp \lz)  + \frac{\hlz}{\gp} (1 - \cos(\gp \lz))\right]\nonumber\\
&\times  \frac{\exp(-\lnmp t)-1}{\anmpsq + \btm^2 + \gp^2}.\label{eq:h}
\end{align}
\end{corollary}
\begin{IEEEproof}
Eq.~\eqref{eq:h} follows by elementary differentiation and integration, cf.~\eqref{eq:def_h}, \eqref{eq:C}.
\end{IEEEproof}
Here, to avoid the obvious difficulties if $\btm = 0$ or $\gp = 0$ in \eqref{eq:h}, we observe that this only happens if $\hly = \hry = 0$ or $\hlz = \hrz = 0$ and in this case
\begin{align}
&\lim\limits_{\btm \to 0} \mathcal{B}_m (\hly \sin(\btm y_0) + \kly \btm \cos(\btm y_0)) (\kly \sin(\btm \ly) + \frac{\hly}{\btm} (1 - \cos(\btm \ly))) =\nonumber\\
&\lim\limits_{\btm \to 0} \frac{1}{\ly \btm^2 \kly^2} \kly \btm \cos(\btm y_0) \kly \sin(\btm \ly) = 1\label{eq:h_limit_const_y}
\end{align}
and, similarly,
\begin{align}
&\lim\limits_{\gp \to 0} \mathcal{C}_p (\hlz \sin(\gp z_0) + \klz \gp \cos(\gp z_0)) (\klz \sin(\gp \lz) + \frac{\hlz}{\gp} (1 - \cos(\gp \lz))) =\nonumber\\
&\lim\limits_{\gp \to 0} \frac{1}{\lz \gp^2 \klz^2} \klz \gp \cos(\gp z_0) \klz \sin(\gp \lz) = 1\label{eq:h_limit_const_z}
\end{align}
and thus, if $\btm = 0$ or $\gp = 0$, we simply replace the corresponding expressions in \eqref{eq:h} with these limits.
This argument could be made more rigorous by considering the constant terms corresponding to $\btm = 0$, $\gp = 0$ separately in \eqref{eq:h}, but, here, we favor a unified treatment over mathematical rigor.
\begin{corollary}
If the boundaries in $y$ and $z$ are reflective, i.e., $\hly = \hry = \hlz = \hrz = 0$, \eqref{eq:h} simplifies to
\begin{equation}
h^r(t) = \sum_{n=1}^{\infty} \mathcal{A}_n^{1,1} C^{1,1}_{n,x}(x_0) \left(\hlx \cos(\alpha^{1,1}_n \lx) - \klx \alpha^{1,1}_n \sin(\alpha^{1,1}_n \lx)\right) \frac{\exp(-D\left(\alpha^{1,1}_n\right)^2 t)-1}{\alpha^{1,1}_n},\label{eq:h_reflective_boundaries}
\end{equation}
where $\lbrace\btm\rbrace$ and $\lbrace\gp\rbrace$ are indexed such that $\beta_1 = \gamma_1 = 0$.

If $\hlx > 0$, $\hdx > 0$, \eqref{eq:h_reflective_boundaries} simplifies further to
\begin{equation}
h^r(t) = \sum_{n=1}^{\infty} \mathcal{A}_n^{1,1} C^{1,1}_{n,x}(x_0) (\hlx \cos(\alpha^{1,1}_n \lx) - \klx \alpha^{1,1}_n \sin(\alpha^{1,1}_n \lx)) \frac{\exp(-D\left(\alpha^{1,1}_n\right)^2 t)}{\alpha^{1,1}_n}\label{eq:hr_wo_constant_terms}.
\end{equation}
\end{corollary}
\begin{IEEEproof}
If $\hly = \hry = \hlz = \hrz = 0$, \eqref{eq:def_btm} and \eqref{eq:def_gp} simplify to $\sin(\beta \ly) = 0$ and $\sin(\gamma \lz) = 0$, respectively, and all terms corresponding to non-zero roots $\btm$, $\gp$ in \eqref{eq:h} vanish.
Then, using the substitutions from \eqref{eq:h_limit_const_y} and \eqref{eq:h_limit_const_z}, \eqref{eq:h_reflective_boundaries} follows directly from \eqref{eq:h}.

Eq.~\eqref{eq:hr_wo_constant_terms} follows from the observation that, if $\hlx > 0$, $\hdx > 0$, all released particles are eventually reuptaken and, hence, $\lim_{t \to \infty} h^r(t) = 0$.
\end{IEEEproof}

Eq.~\eqref{eq:h_reflective_boundaries} is exactly the result that was obtained for the one-dimensional model in \cite{lotter20a}.

Since \eqref{eq:hr_wo_constant_terms} is a weighted sum of decaying exponentials, we expect that $h^r$ is dominated by the exponential with the slowest decay for large values of $t$.
Thus, we obtain the following approximation for large $t$:
\begin{equation}
h^r(t) \approx \mathcal{A}_1^{1,1} C^{1,1}_{1,x}(x_0) \left(\hlx \cos(\alpha^{1,1}_1 \lx) - \klx \alpha^{1,1}_1 \sin(\alpha^{1,1}_1 \lx)\right) \frac{\exp(-D\left(\alpha^{1,1}_1\right)^2 t)}{\alpha^{1,1}_1}.\label{eq:hr_first_term}
\end{equation}
The accuracy of this approximation will be investigated in Section \ref{sec:results}, cf.~Fig.~\ref{fig:reuptake_ht}. 

For non-reflective boundaries, $h(t)$ is asymptotically still dominated by the first summand in \eqref{eq:h}, and the first-term approximation
\begin{align}
h(t) &\approx \mathcal{A}_1^{1,1} \mathcal{B}_1 \mathcal{C}_1 C_0(x_0,y_0,z_0)\left[\hlx \alpha^{1,1}_1 \cos(\alpha^{1,1}_1 \lx) - \klx \left(\alpha^{1,1}_1\right)^2 \sin(\alpha^{1,1}_1 \lx)\right] \nonumber\\
&\times \left[\kly \sin(\beta_1 \ly) + \frac{\hly}{\beta_1} (1 - \cos(\beta_1 \ly))\right]\left[\klz \sin(\gamma_1 \lz)  + \frac{\hlz}{\gamma_1} (1 - \cos(\gamma_1 \lz))\right]\nonumber\\
&\times  \frac{\exp(-\lambda^{1,1}_1 t)}{\left(\alpha^{1,1}_1\right)^2 + \beta_1^2 + \gamma_1^2}\label{eq:def_h_first_term}
\end{align}
provides an accurate estimate of the \ac{CIR} decay, cf.~Fig.~\ref{fig:glial_uptake}.

\begin{remark}
Comparing the impact of glial cell uptake on \eqref{eq:def_h_first_term}, represented mainly by $\beta_1$ and $\gamma_1$, with the effect of enzymatic degradation on the molecule concentration as studied in \cite{noel14}, we note that the effects of uptake and enzymatic degradation are formally similar in that both provide an additional multiplicative exponential decay term in the respective expressions.
However, as we will see in Section \ref{sec:results:glial_cell_uptake}, the effect of glial cell uptake may, in some situations, act more selectively on the undesired tail of the signal.
\end{remark}

This observation concludes the derivation of the \ac{CIR}.
\subsubsection{Number of Reuptaken Particles}

Now, we consider the number of particles taken up at the presynaptic neuron and at the glial cell.
\begin{corollary}
The number of reuptaken particles at the presynaptic neuron as defined in \eqref{eq:def_i} is given as follows
\begin{align}
i(t) &= \hlx \sum_{n=1}^{\infty}\sum_{m=1}^{\infty}\sum_{p=1}^{\infty} \anmp \mathcal{A}_n^{m,p} \mathcal{B}_m \mathcal{C}_p C_0(x_0,y_0,z_0) \left[\kly \sin(\btm \ly) + \frac{\hly}{\btm} \left(1 - \cos(\btm \ly)\right)\right]\nonumber\\
&\times \left[\klz \sin(\gp \lz)  + \frac{\hlz}{\gp} \left(1 - \cos(\gp \lz)\right)\right] \frac{1-\exp(-\lnmp t)}{\anmpsq + \btm^2 + \gp^2}.\label{eq:i}
\end{align}
The number of particles uptaken at the glial cell, $i_{-}(t)$ as defined in \eqref{eq:def_ig} is given by
\begin{align}
i_{-}(t) &= \sum_{n=1}^{\infty}\sum_{m=1}^{\infty}\sum_{p=1}^{\infty} \mathcal{A}_n^{m,p} \mathcal{B}_m \mathcal{C}_p C_0(x_0,y_0,z_0) \left[D \sin(\anmp \lx) + \frac{\hlx}{\anmp} \left(1 - \cos(\anmp \lx)\right)\right]\nonumber\\
&\times \left\lbrace\left[\hly C_{m,y}(0) + \hry C_{m,y}(\ly)\right] \left[\klz \sin(\gp \lz)  + \frac{\hlz}{\gp} \left(1 - \cos(\gp \lz)\right)\right]\right.\nonumber\\
&\left.+\left[\kly \sin(\btm \ly)  + \frac{\hly}{\btm} \left(1 - \cos(\btm \ly)\right)\right] \left[\hlz C_{p,z}(0) + \hrz C_{p,z}(\lz)\right]\right\rbrace\nonumber\\
&\times\frac{1-\exp(-\lnmp t)}{\lnmp}.\label{eq:ig}
\end{align}
\end{corollary}
\begin{IEEEproof}
We plug \eqref{eq:pab_lb} into \eqref{eq:def_i} and apply elementary integration to \eqref{eq:C} to obtain \eqref{eq:i}.
Similarly, we use \eqref{eq:rb_l} and \eqref{eq:rb_r} to obtain \eqref{eq:ig}.
\end{IEEEproof}
\begin{corollary}
If the boundaries in $y$ and $z$ are reflective, i.e., $\hly = \hry = \hlz = \hrz = 0$, \eqref{eq:i} simplifies to   
\begin{align}
i^r(t) &= \hlx \sum_{n=1}^{\infty} \mathcal{A}_n^{1,1} C^{1,1}_{n,x}(x_0) \frac{1-\exp(-D\left(\alpha^{1,1}_n\right)^2 t)}{\alpha^{1,1}_n},\label{eq:i_reflective_boundaries}
\end{align}
where, again, we have set $\beta_1 = \gamma_1 = 0$.
If $\hdx > 0$, \eqref{eq:i_reflective_boundaries} simplifies to 
\begin{equation}
i^r(t) = 1 - \hlx \sum_{n=1}^{\infty} \mathcal{A}_n^{1,1} C^{1,1}_{n,x}(x_0) \frac{\exp(-D\left(\alpha^{1,1}_n\right)^2 t)}{\alpha^{1,1}_n}.\label{eq:ir_wo_constant_terms}
\end{equation}
\end{corollary}
\begin{IEEEproof}
We use the substitutions from \eqref{eq:h_limit_const_y} and \eqref{eq:h_limit_const_z} to obtain \eqref{eq:i_reflective_boundaries} from \eqref{eq:i}.
If $\hdx > 0$, we can further exploit \eqref{eq:i_cdf} to obtain $1 = \lim\limits_{t \to \infty} i^r(t) = \hlx \sum_{n=1}^{\infty} \mathcal{A}_n^{1,1} C^{1,1}_{n,x}(x_0) / \alpha^{1,1}_n$ and \eqref{eq:ir_wo_constant_terms} follows.
\end{IEEEproof}

Hence, we note that, if the boundaries in $y$ and $z$ are reflective, the three-dimensional domain $\Omega$ is in terms of $h(t)$ and $i(t)$ equivalent to the one-dimensional domain $[0;\lx]$.

From \eqref{eq:ir_wo_constant_terms}, we obtain
\begin{equation}
i^r(t) \approx 1 - \hlx  C^{1,1}_{1,x}(x_0) \mathcal{A}_1^{1,1} \frac{\exp(-D\left(\alpha^{1,1}_1\right)^2 t)}{\alpha^{1,1}_1} =: i^r_1(t).\label{eq:ir_first_term}
\end{equation}
as an approximation for the number of particles taken up at the presynaptic neuron.
The accuracy of this approximation is investigated in Section \ref{sec:results}, cf.~Fig.~\ref{fig:reuptake_it}.

\section{CIR Decay Rate}
\label{sec:cir_decay_rate}
From the observations made in the previous section, we conclude that the long-term system behavior is governed by the values of $\alpha^{1,1}_1$, $\beta_1$, and $\gamma_1$. Now, we characterize these in order to develop a better understanding of the limits and mechanisms that determine the postsynaptic signal as well as the reuptake of molecules. To this end, we start by characterizing $\alpha^{1,1}_1$ for the special case of reflective boundaries in $y$ and $z$, and then discuss the more general case of partially absorbing boundaries.

\subsection{Reflective Boundaries}\label{sec:cir_decay_rate:reflective_case}
From \eqref{eq:hr_first_term} and \eqref{eq:ir_first_term}, we observe that, if the boundaries in $y$ and $z$ are reflective, the smallest root of \eqref{eq:def_an}, $\alpha^{1,1}_1$, dictates the decay rate of the \ac{CIR}, and, at the same time, governs the rate of the presynaptic reuptake.
For ease of notation, we drop the dependence of $\alpha^{1,1}_1$ on $\beta_1$ and $\gamma_1$ and write $\alpha_1$ instead.

Now, to characterize $\alpha_1$, we consider both sides of \eqref{eq:def_an}, which for $\hly = \hry = \hlz = \hrz = 0$ simplify to
\begin{equation}
\tan(\alpha l_x) = \frac{\alpha^2 D(\hlx+\hrx)-\hlx\hdx}{\alpha\left\lbrack\alpha^2 D^2-\left(\hlx\hrx+\hdx D\right)\right\rbrack} =: u(\alpha).\label{eq:def_an_1d}
\end{equation}
All solutions of this equation are real and we seek the smallest positive one.

\begin{proposition}\label{prop:alpha_1_bounds}
Let $t_{\lx} = \frac{\lx^2}{2 D}$ denote the time that a particle needs (on average) to travel distance $\lx$ if it diffuses with diffusion coefficient $D$.
Then, the following inequality holds for the rate which determines the exponential decay in \eqref{eq:hr_first_term} and \eqref{eq:ir_first_term}, $D \alpha_1^2$:
\begin{equation}
D \alpha_1^2 < \min\left\lbrace\frac{\pi^2}{8}\frac{1}{t_{\lx}},\hdx \frac{\hlx}{\hlx+\hrx}\right\rbrace\label{eq:alpha_1_bounds}
\end{equation}
\end{proposition}
\begin{IEEEproof}
Let us first examine the middle term in \eqref{eq:def_an_1d}, which we denote by $u(\alpha)$.
It has two non-negative poles, one at $\alpha=0$ and one at $\alpha=\sqrt{\hlx\hrx+\hdx D}/D=:\alpha_{(2)}$.
It can easily be shown that $\lim_{\alpha \to 0^+} u(\alpha) = + \infty$ and $\lim_{\alpha \to \alpha_{(2)}^-} u(\alpha) = - \infty$.
Furthermore, $u(\alpha)$ possesses exactly one positive root at $\alpha=\sqrt{\frac{\hlx\hdx}{D(\hlx+\hrx)}}$.
The left-hand side of \eqref{eq:def_an_1d} is periodic with periodicity $\frac{\pi}{\lx}$ and has poles at $\frac{\pi}{2\lx}+\frac{k\pi}{\lx}$ and roots at $\frac{k\pi}{\lx}$, $k \in \mathbb{Z}$.
Within each interval $(\frac{k\pi}{\lx}-\frac{\pi}{2\lx};\frac{k\pi}{\lx}+\frac{\pi}{2\lx})$, $\tan(\alpha l_x)$ is strictly monotonically increasing and continuous and thus takes on all values from $-\infty$ to $+\infty$.
Hence, we conclude that the smallest root of \eqref{eq:def_an_1d}, $\alpha_1$,
\begin{enumerate*}[label=(\roman*)]
    \item is contained in the interval $(0;\frac{\pi}{2 \lx})$,\label{alp:prop:1}
    \item is smaller than $\sqrt{\frac{\hlx\hdx}{D(\hlx+\hrx)}}$.\label{alp:prop:2}
\end{enumerate*}
Now, from \ref{alp:prop:1}, we obtain $D \alpha_1^2 < \frac{D \pi^2}{4 \lx^2} = \frac{\pi^2}{8}\frac{1}{t_{\lx}}$, and \ref{alp:prop:2}, on the other hand, yields $D \alpha_1^2 < \hdx \frac{\hlx}{\hlx+\hrx}$.
This completes the proof.
\end{IEEEproof}

Thus, because $\frac{\hlx}{\hlx+\hrx} < 1$ the decay of $h^r(t)$ can never be faster than the rate of the desorption reaction at the postsynapse, $\hdx$.
On the other hand, channel clearance is fast, if presynaptic reuptake dominates postsynaptic binding, i.e., $\hlx \gg \hrx$, and slow, if molecules tend to bind to the postsynapse instead of being reuptaken, i.e., $\hrx \gg \hlx$.

From Proposition~\ref{prop:alpha_1_bounds}, we see that the decay of the \ac{CIR} can either be limited by the time that particles need to cross the synapse or by the reaction kinetics at the pre- and postsynapse, depending on whether $\frac{\pi^2}{8}\frac{1}{t_{\lx}} < \hdx \frac{\hlx}{\hlx+\hrx}$ or $\hdx \frac{\hlx}{\hlx+\hrx} < \frac{\pi^2}{8}\frac{1}{t_{\lx}}$.
We call the regime $\frac{\pi^2}{8}\frac{1}{t_{\lx}} < \hdx \frac{\hlx}{\hlx+\hrx}$ {\em diffusion-limited (in $x$)}, and the regime $\hdx \frac{\hlx}{\hlx+\hrx} < \frac{\pi^2}{8}\frac{1}{t_{\lx}}$ {\em reaction-limited (in $x$)}.
Thus, if the two neurons are very close, i.e., $\lx$ is small, synaptic transmission happens in the reaction-limited regime, while fast molecule dissociation at the postsynapse and fast presynaptic uptake favor the diffusion-limited regime.
Proposition~\ref{prop:alpha_1_bounds} reveals that both, diffusion and reaction, impose fundamental bounds on signal decay and particle reuptake and, more importantly, these bounds are independent of each other, i.e., in the reaction-limited regime, independent of the geometry of the synaptic cleft, signal decay and reuptake can never exceed $\frac{\hdx \hlx}{\hlx+\hrx}$, and, similarly, in the diffusion-limited regime, even for arbitrarily fast reactions, $\frac{\pi^2}{8 t_{\lx}}$ bounds the maximum achievable signal decay.

So far, we have seen how the system parameters influence the (un)achievable range of $\alpha_1$.
Now, we examine their direct impact on the value of $\alpha_1$.
To this end, we fix any $\alpha^* \in (0;\frac{\pi}{2 \lx})$ and consider $u$ instead as a function of any channel parameter which is currently under inspection, e.g.~$u(\hlx)$.
Similarly, we consider $\alpha_1 = \alpha_1(\hlx)$ as a function of that parameter.
Finally, for the investigation of $\lx$, we consider $\tan(\alpha \lx)$ as a function of $\lx$.
\begin{proposition}\label{prop:alpha_1_derivatives}
$\alpha_1$ is increasing in $\hlx$ and $\hdx$ and decreasing in $\hrx$ and $\lx$, i.e.,
\begin{align}
\frac{\textrm{d}\alpha_1}{\textrm{d} \hlx},\frac{\textrm{d}\alpha_1}{\textrm{d} \hdx} &> 0, &\frac{\textrm{d}\alpha_1}{\textrm{d} \hrx},\frac{\textrm{d}\alpha_1}{\textrm{d} \lx} &< 0.\label{eq:da_dhrx}
\end{align}    
\end{proposition}
\begin{IEEEproof}
Due to the strict monotonicity of $\tan(\cdot)$, $\frac{\textrm{d}u(\hlx)}{\textrm{d} \hlx} > 0$ implies $\frac{\textrm{d}\alpha_1}{\textrm{d} \hlx} > 0$.
The same holds for $\hrx$ and $\hdx$, and for $\lx$, $\frac{\textrm{d}\tan(\alpha^* \lx)}{\textrm{d} \lx} > 0$ implies $\frac{\textrm{d}\alpha_1}{\textrm{d} \lx} < 0$.
By simple differentiation, we obtain $\frac{\textrm{d}u(\hlx)}{\textrm{d} \hlx},\frac{\textrm{d}u(\hdx)}{\textrm{d} \hdx},\frac{\textrm{d}\tan(\alpha^* \lx)}{\textrm{d} \lx} > 0$, $\frac{\textrm{d}u(\hrx)}{\textrm{d} \hrx} < 0$ and, hence, \eqref{eq:da_dhrx} follows.
This completes the proof.
\end{IEEEproof}

Thus, increasing reuptake and desorption accelerates channel clearance, while larger postsynaptic adsorption and synaptic cleft width slow it down.

\subsection{Non-Reflective Boundaries}
In the general case of partially absorbing boundaries in $y$ and $z$, the decay of the \ac{CIR} is governed by the three-dimensional decay rate $\lambda^{1,1}_1 = D\left\lbrack\left(\alpha^{1,1}_1\right)^2+\beta_1^2+\gamma_1^2\right\rbrack$, see \eqref{eq:def_h_first_term}.

Let us first consider $\alpha^{1,1}_1$. 
If $\alpha^{1,1}_1$ is real, the discussion parallels the discussion of $\alpha_1$ with minor modifications due to $\beta_1$ and $\gamma_1$.

\begin{proposition}
If $\alpha^{1,1}_1 \in \mathbb{R}$, the exponential decay rate $\lambda^{1,1}_1$ of the first-term approximation of $h(t)$ as defined in \eqref{eq:def_h_first_term} is bounded by the following inequality:
\begin{equation}
\lambda^{1,1}_1 < \min\left\lbrace\frac{\pi^2}{8}\frac{1}{t_{\lx}} + D (\beta_1^2+\gamma_1^2),\hdx\frac{\hlx}{\hlx+\hrx}\right\rbrace
\end{equation}
\end{proposition}
\begin{IEEEproof}
With similar arguments as in the proof of Proposition \ref{prop:alpha_1_bounds}, one can show that, if $\alpha^{1,1}_1 \in \mathbb{R}$, $\alpha^{1,1}_1$,
\begin{enumerate*}[(\roman*)]
    \item is contained in the interval $(0;\frac{\pi}{2 \lx})$,\label{lbd:prop:1}
    \item is smaller than $\sqrt{\frac{\hlx\hdx}{D(\hlx+\hrx)}-(\beta_1^2+\gamma_1^2)}$.\label{lbd:prop:2}
\end{enumerate*}
Hence, we conclude that for non-reflective boundaries the upper bound on the maximum decay rate induced by the channel geometry is
\begin{equation}
\lambda^{1,1}_1 < \frac{\pi^2}{8}\frac{1}{t_{\lx}} + D (\beta_1^2+\gamma_1^2),\label{eq:ub_geom_gen}
\end{equation}
and the reaction rate induced upper bound is the same as in the one-dimensional scenario, namely
\begin{equation}
\lambda^{1,1}_1 < \hdx\frac{\hlx}{\hlx+\hrx}.
\end{equation}
This completes the proof.
\end{IEEEproof}

For many practical scenarios, $\beta_1$ and $\gamma_1$ are too small relative to $\pi^2/ \left(8 t_{\lx}\right)$ so as to significantly increase the bound in \eqref{eq:ub_geom_gen}.
This means that for such scenarios the maximum \ac{CIR} decay of a synaptic transmission system with glial cell uptake can also be achieved by an equivalent one-dimensional system with high presynaptic reuptake and postsynaptic desorption rate.
The exception to this rule are synaptic transmission systems in which the channel is ``wider than high'', i.e., $\ly$ and $\lz$ are relatively small compared to $\lx$.
Such systems can for example be found in the retina \cite{heidelberger05}.
In this case, as the values of $\beta_1$ and $\gamma_1$ scale with the reciprocal of $\ly$ and $\lz$, respectively, $\beta_1$ and $\gamma_1$ are in a range in which $\pi^2/ \left(8 t_{\lx}\right) + D (\beta_1^2+\gamma_1^2) \gg \pi^2/ \left(8 t_{\lx}\right)$.

Now, let us consider the case $\alpha^{1,1}_1 \not\in \mathbb{R}$.
In order for \eqref{eq:def_an} to allow for an imaginary root, two conditions must be fulfilled.
First, there must be a flow of molecules back from the postsynaptic boundary into the system, i.e., $\hdx > 0$, and, second, absorption in $y$ and/or $z$ must be strong, i.e., $\beta_m^2+\gamma_p^2$ must be large.
Under these conditions, a pair of complex conjugate, purely imaginary, roots emerges as solution of \eqref{eq:def_an}.
Let us write $\alpha^{1,1}_1 = j \theta$.
Then, we have $\lambda^{1,1}_1 = D\left(-\theta^2+\beta_1^2+\gamma_1^2\right)$ and see that $\alpha^{1,1}_1$ contributes {\em negatively} to the decay of the \ac{CIR}, i.e., particle transmission in $x$ direction {\em slows down} the signal decay.

At first glance, this seems to be counter-intuitive, but actually it is a very sensible insight: First, being adsorbed to the postsynaptic membrane prevents a particle from being absorbed by the ``hungry'' glial cell and, second, a particle that has desorbed from the postsynaptic membrane may re-adsorb again and in this way contribute positively to the postsynaptic signal.
This is true in general, but when $\beta_m^2+\gamma_p^2$ is large, uptake at glial cells is so fast that desorption from the postsynaptic membrane acts effectively as a molecule source.
The signal decay in the synaptic cleft is then clearly dictated by the glial cell.

Now, to complete our discussion of $\lambda^{1,1}_1$, let us examine $\beta_1^2$, the same analysis holds for $\gamma_1^2$.
We have already seen that if $\hly=\hry=0$, $\beta_1=0$, i.e., there is zero contribution from $\beta_1$ to $\lambda^{1,1}_1$.
Thus, consider the case in which at least one of $\hly$ and $\hry$ is non-zero.
Then, $\beta_1^2$ is defined as the smallest {\em non-zero} solution of \eqref{eq:def_btm}.

\begin{proposition}\label{prop:beta_1_bounds}
If $\hly + \hry > 0$, the following inequalities hold:
\begin{align}
\beta_{(1)} < \beta_1 < \frac{\pi}{2 \ly} &\textrm{ if } \beta_{(1)} < \frac{\pi}{2 \ly}\label{eq:def_bm_reac_lim_reg_a} \\
\frac{\pi}{2 \ly} < \beta_1 \leq \frac{\pi}{\ly} &\textrm{ if } \beta_{(1)} > \frac{\pi}{\ly}\label{eq:def_bm_diff_lim_reg} \\
\frac{\pi}{2 \ly} < \beta_1 < \beta_{(1)} &\textrm{ otherwise},\label{eq:def_bm_reac_lim_reg_b}
\end{align}
where $\beta_{(1)}=\sqrt{\frac{\hly \hry}{D^2}}$ and equality in \eqref{eq:def_bm_diff_lim_reg} is attained if and only if $\kly=\kry=0$.
\end{proposition}
\begin{IEEEproof}
The analysis here is a bit different from the above analysis of $\alpha_1$.
First, we consider a partially adsorbing boundary, $\kly,\kry = D \neq 0$.
We note that the right-hand side of \eqref{eq:def_btm}, which we denote by $v(\beta) = \frac{\beta D(\hly+\hry)}{\beta^2 D^2-\hly\hry}$, has a root at $\beta=0$ and a pole at $\beta_{(1)}=\sqrt{\frac{\hly \hry}{D^2}}$.
Furthermore, $\lim_{\beta \to \beta^-_{(1)}} v(\beta) = -\infty$ and $v(\beta) < 0, \forall \beta \in (0;\beta_{(1)})$.
As the left-hand side of \eqref{eq:def_btm} is periodic with periodicity $\frac{\pi}{\ly}$, has poles at $\frac{\pi}{2\ly}+\frac{k\pi}{\ly}$, and is strictly monotonically increasing in each interval $\left(\frac{\pi}{2\ly}+\frac{k\pi}{\ly};\frac{\pi}{2\ly}+\frac{(k+1)\pi}{\ly}\right)$, $k \in \mathbb{Z}$, inequalities \labelcref{eq:def_bm_reac_lim_reg_a,eq:def_bm_diff_lim_reg,eq:def_bm_reac_lim_reg_b} follow.
Now, if $\kly=\kry=0$, \eqref{eq:def_btm} simplifies to $\sin(\beta \ly) = 0$, and thus, $\beta_m = m \frac{\pi}{\ly}$, $m \in \mathbb{N}$.
Hence, we have $\beta_1 = \frac{\pi}{\ly}$ and the upper bound in \labelcref{eq:def_bm_diff_lim_reg} is attained.
This completes the proof.
\end{IEEEproof}

To simplify the interpretation of this result, let us consider the special case $\hly=\hry=\kappa_y$.

Then, $\beta_{(1)} > \frac{\pi}{\ly}$ if and only if $\kappa_y > \pi \frac{\ly}{t_{\ly}}$, where $t_{\ly} = \frac{\ly^2}{2 D}$ denotes the time that a particle needs (on average) to travel a distance $\ly$ with diffusion coefficient $D$.
Thus, if the reaction kinetics at the boundaries are sufficiently fast relative to the particle velocity, \eqref{eq:def_bm_diff_lim_reg} holds, and $\beta_1$ is constrained by the channel geometry only.
Similar to before, we call this regime {\em diffusion-limited (in $y$)}, but, here, the particle velocity (relative to the channel length in $y$) provides not only an upper, but also a lower bound for the contribution of particle loss in $y$ to the overall decay rate.
This implies that, if $\ly$ decreases, e.g.~due to a change in the morphology of a glial cell, as long as reactions at the glial cell happen fast enough, i.e., $\beta_{(1)} > \frac{\pi}{\ly}$ holds, we have a stronger guarantee on the minimum signal decay.
For fixed reaction rates, however, if $\ly$ grows beyond limits, \eqref{eq:def_bm_diff_lim_reg} holds, $\beta_1$ approaches $0$, and we recover the reflective scenario.

On the other hand, if the reaction kinetics are relatively slow, i.e., $\kappa_y < \pi \frac{\ly}{t_{\ly}}$, and either of \eqref{eq:def_bm_reac_lim_reg_a} or \eqref{eq:def_bm_reac_lim_reg_b} applies, $\beta_1$ is bounded by the reaction kinetics, either from below or from above.
We call this regime again {\em reaction-limited (in $y$)}, though we note that in $x$ reaction kinetics exclusively provided an {\em upper bound}.

The next proposition concludes this analysis with results on the dependence of $\beta_1$ on $\hly$, $\hry$, and $\ly$.
\begin{proposition}
$\beta_1$ as a function of $\hly$, $\hry$, and $\ly$ is increasing in $\hly$ and $\hry$ and decreasing in $\ly$, i.e.,
\begin{align}
\frac{\textrm{d}\beta_1}{\textrm{d} \hly},\frac{\textrm{d}\beta_1}{\textrm{d} \hry} &> 0, \qquad \frac{\textrm{d}\beta_1}{\textrm{d} \ly} < 0.\label{eq:db_dly}
\end{align}    
\end{proposition}
\begin{IEEEproof}
Similar to the proof of Proposition \ref{prop:alpha_1_derivatives}, \eqref{eq:db_dly} follows from the monotonicity of $\tan(\beta \ly)$ and elementary differentiation of $v(\beta)$ as defined in the proof of Proposition \ref{prop:beta_1_bounds}.
\end{IEEEproof}

\section{Particle-based Simulation}
\label{sec:pb_sim}
To verify the accuracy of the analytical expressions derived in Section~\ref{sec:cir}, three-dimensional {\em stochastic} particle-based computer simulations were conducted.
To this end, we adopted the simulator design from \cite{andrews09}.
The simulator was implemented in the programming language C.
Simulations were conducted on a computing cluster comprised of several machines that executed the simulation program in parallel.
To ensure the reproducibility of our results, all random seeds as well as the program version of the simulator were serialized and archived together with the data.
After simulation, the results from the different simulation runs were aggregated and averaged.
\subsection{Simulator Design}\label{sec:pb_sim:sim_design}
Brownian particle motion is simulated by updating the position of each point-like particle at each time step with a three-dimensional jointly independent Gaussian random vector $[X, Y, Z] \sim \mathcal{N}(\mathbf{0}_{1 \times 3},\sigma^2 \mathbf{I}_{3 \times 3})$.
Here, $\mathcal{N}(\boldsymbol\mu,\boldsymbol{\mathcal{C}})$ denotes the multivariate Gaussian distribution with mean vector $\boldsymbol\mu$ and covariance matrix $\boldsymbol{\mathcal{C}}$, and $\mathbf{0}_{1 \times M}$ and $\mathbf{I}_{M \times M}$ denote the $1 \times M$ all-zero vector and the $M \times M$ identity matrix, respectively.
The variance $\sigma^2$ is a function of the particle diffusion coefficient $D$ and the simulation time step $\Delta t$, $\sigma^2 = 2 D \Delta t$.
Accordingly, the {\em root mean step length (rms)} of a simulated particle is given by
\begin{equation}
s = \sqrt{2 D \Delta t}.\label{eq:def_rms}
\end{equation}

In the simulation, the probability that a particle is adsorbed when hitting a homogeneous boundary is computed from the respective adsorption coefficient using \cite[eq.~(21)]{andrews09}.
If a particle hits a boundary and is not absorbed, it is reflected using ballistic reflection.
Particles hitting a receptor at the postsynaptic boundary are absorbed with probabilities computed as in \cite[eqs.~(37), (32)]{andrews09} from the intrinsic binding rate of molecules to receptors, $\kad$, and the intrinsic desorption rate constant, $\hdx$.
Desorbed particles are placed apart from the boundary according to \cite[eq.~(35)]{andrews09}.

\subsection{Boundary Homogenization}

In order to compare simulation data and analytical results, boundary homogenization at the postsynaptic boundary is performed, i.e., the heterogeneous surface covered by individual receptors as simulated in the particle-based simulator is identified with an equivalent homogeneous surface as modeled in \eqref{eq:pab_rb}.
For {\em irreversible} molecule binding to individual receptors, this technique is well-developed \cite{zwanzig1991,berezhkovskii04}.
However, for {\em reversible} reactions, to the best of the authors' knowledge, it is not available yet.
Indeed, the existing general theory for reversible diffusion-influenced reactions is rather inaccessible for the non-expert user \cite{gopich02,berezhkovskii13a}.
Here, there are two main reasons why existing results for {\em irreversible} ligand-receptor binding, such as \cite{berezhkovskii04}, cannot be applied.

First, directly after unbinding from a receptor, molecules are physically still close to this receptor and hence, the concentration of molecules at the receptor-covered surface is non-uniform.
This implies that the probability of hitting again a receptor when hitting the surface is higher for molecules that have just desorbed than for molecules which were never bound to a receptor.
In other words, reversibly binding molecules have a (short-term) memory, a state, which impacts their binding properties.
This observation is in contrast to memoryless irreversible binding.
We can, however, circumvent this problem in a biologically plausible manner \cite{zarnitsyna07} by making the assumption that there is a (short-term) mechanism which prevents ligands from immediate rebinding to the same receptor.
Thus, we choose $\Delta t$ and the receptor radius $r$ such that $s$ is larger than $r$.
As the displacement of particles after desorption scales with $s$ (see \cite[eq.~(35)]{andrews09}), this has the effect that desorbed particles are more likely to see a representative part of the boundary before possibly adsorbing to a receptor again.
This is in fact equivalent to blocking the desorbed particle for some time from rebinding and, hence, renders molecule binding a memoryless process.

Second, boundary homogenization for irreversible reactions is performed such that the {\em average particle lifetimes} in presence of the receptor-covered surface and the homogeneous surface, respectively, are identical \cite{berezhkovskii04}.
This requirement, however, is not meaningful for reversible reactions because generally, in the absence of any irreversible molecule degradation mechanism, the lifetime of molecules in the presence of a reversibly absorbing surface is infinite.
Thus, we compute the effective adsorption coefficient for the homogenized boundary, $\hrx$, from $\kad$ using computer-assisted boundary homogenization similar to \cite{berezhkovskii04}.
Unlike \cite{berezhkovskii04}, however, we require analytical and numerical results to produce the same {\em average particle lifetime}, instead of the same {\em steady-state concentrations}.
To this end, we conduct particle-based simulations in $\Omega$ \eqref{eq:domain} without presynaptic reuptake and with reflective boundaries in $y$ and $z$ until the system approaches its steady-state.
Next, we use our analytical result in \eqref{eq:C} to compute the steady-state number of particles adsorbed to the postsynaptic boundary for this scenario, $h^{\infty} = 1 - \lx \hdx / \left(\hrx + \lx \hdx\right) = \hrx / \left(\hrx + \lx \hdx\right)$.
Now, we denote the simulated steady-state number of adsorbed particles as $h^{\infty}_{\textrm{sim}}$ and require $h^{\infty}_{\textrm{sim}} = h^{\infty}$, which we solve for $\hrx$ to obtain $\hrx = \lx \hdx h^{\infty}_{\textrm{sim}}/(1-h^{\infty}_{\textrm{sim}})$.
Next, we compute $h^{\infty}_{\textrm{sim}}$ for many different parameter sets and find in this way that
\begin{equation}
\hrx = 0.984 \rho \kad,\label{eq:boundary_homogenization}
\end{equation}
where $\rho$ denotes the fraction of the postsynaptic surface covered by receptors, provides an excellent approximation for the range of parameters that we consider (see Section \ref{sec:results}, cf.~Table~\ref{tab:sim_params}).

In contrast to \cite{berezhkovskii04} and earlier results, \eqref{eq:boundary_homogenization} indicates that, for reversible binding, $\hrx$ does not vary with the receptor radius $r$, as long as $\rho$ is kept fixed.
This is intuitive; it is known that the receptor radius impacts the average particle lifetime (given a constant surface coverage) \cite{berezhkovskii04}.
However, this is a transient phenomenon and does not apply to the steady-state concentration considered here.

\section{Numerical Results}
\label{sec:results}
In this section, we evaluate the analytical expressions for the \ac{CIR} and the number of reuptaken particles derived in Section \ref{sec:cir} and compare them with results from particle-based simulation as outlined in Section \ref{sec:pb_sim}.
For the computation of the analytical results, the infinite sums in \eqref{eq:h}, \eqref{eq:hr_wo_constant_terms}, \eqref{eq:i}, \eqref{eq:ig}, and \eqref{eq:ir_wo_constant_terms} were truncated after $100$ terms in $x$ and $20$ terms in $y$ and $z$, respectively, i.e., $n \in \{1,\ldots,100\}$, $m \in \{1,\ldots,20\}$, and $p \in \{1,\ldots,20\}$.
Furthermore, all analytical results were scaled with the number of particles released in the particle-based simulations, $N$.
The results from particle-based simulation were averaged over $50$ realizations.
Unless indicated otherwise, particles were released at $(x_0,y_0,z_0) = (s,\ly/2,\lz/2)$, where $s$ is defined in \eqref{eq:def_rms}, i.e., particle release is centered in the $y$-$z$-plane and occurs one average simulation step length from the presynaptic boundary in $x$.
Particle release at $x_0 = 0.0$ would in principle also be covered by our analytical results.
However, we chose $x_0 = s$ because after release in real synapses, \acp{NT} are located {\em in front of} the presynaptic membrane and not {\em on} the presynaptic membrane \cite{ashery14}.

The remainder of this section is organized as follows.
In Subsection \ref{sec:results:parameters}, we provide details on the choice of the default parameters.
In Subsection~\ref{sec:results:model_comparison}, we compare the results obtained with our model with previous experimental and computational studies.
Next, we compare in Subsection \ref{sec:results:reuptake} the impact of presynaptic reuptake and postsynaptic desorption on the \ac{CIR} and on the number of reuptaken particles, respectively, {\em without} uptake at the glial cell.
In Subsection \ref{sec:results:glial_cell_uptake}, molecule uptake at the glial cell subject to different glial cell morphologies and different \ac{NT} transporter densities is considered.
Finally, we conclude this section investigating the impact of the molecule release location on the \ac{CIR} in the presence of spillover in Subsection \ref{sec:results:release_location}.

\subsection{Choice of Default Parameters}\label{sec:results:parameters}
Whenever available, sensible default values for the model parameters were adopted from experimental studies of glutamatergic excitatory hippocampal synapses as indicated in Table \ref{tab:sim_params}.
Some parameters that were not readily available were determined as follows.
The binding and unbinding rates of our simplified two-state postsynaptic reaction scheme, $\hrx$ and $\hdx$, were computed from the binding rates to individual postsynaptic receptors \cite{holmes95} and from the postsynaptic receptor density reported in \cite{jonas93}.
The glial cell uptake rate constants $\hly$, $\hry$, $\hlz$, and $\hrz$ were obtained from the binding rate constant to an individual \ac{NT} transporter and the transporter density at glial cells reported in \cite{wadiche98} and \cite{lehre98}, respectively.
In lack of reliable experimental values for the presynaptic reuptake rate constant, we assumed in accordance with \cite{holmes95} a three-fold larger ``uptake capacity'' at glial cells compared to neurons.
As the ``uptake capacity'' is defined as the product of transport rate per individual transporter, surface to volume ratio, and transporter density at the particular membrane, assuming that the transport rate is the same at glial cells and neurons, we compute that the transporter density at glial cells is approximately $20$ times larger than at neural cells and in this way obtain $\hlx = \hly/20 = 1.3 \times 10^{-6}$~$\si{\micro\meter\per\micro\second}$.

Finally, the simulation time step, $\Delta t$, the radius of the model receptors, $r$, and the postsynaptic receptor coverage, $\rho$, were chosen such that \ref{ass:4} from Section \ref{sec:model} is fulfilled, i.e., such that boundary homogenization can be applied.

In the default parameter setting, $\alpha^{1,1}_1$ is imaginary and $\beta_1$ and $\gamma_1$ are close to the reaction-induced lower bound from \eqref{eq:def_bm_reac_lim_reg_a}.
Hence, the system is dominated by the reaction kinetics at the glial cell.

\subsection{Model Comparison with Previous Studies}\label{sec:results:model_comparison}

\begin{figure}
    \centering
    \includegraphics[width=.5\linewidth]{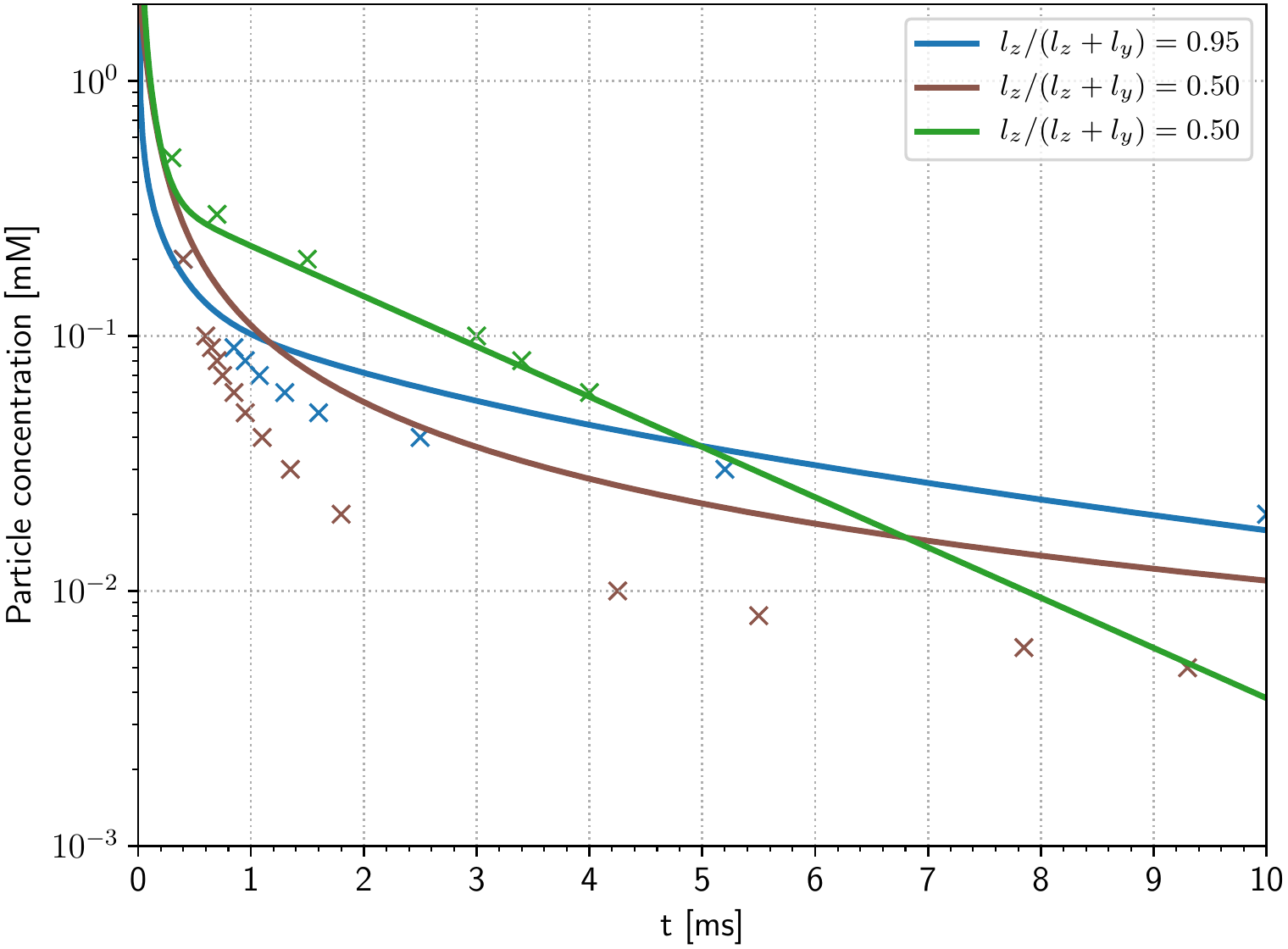}
    \vspace{-4mm}
    \caption{Curves: Concentration of \acp{NT} at the cleft center in \si{\milli M} as predicted by Proposition 1. Please refer to the text for the respective parameter settings. Markers: Data from \cite{rusakov01} (brown and blue) and \cite{clements96} (green).}
    \label{fig:rusakov_clements}
    \vspace{-8mm}
\end{figure}

First, we compare Proposition~\ref{prop:C} with the synaptic \ac{NT} concentration obtained in \cite{rusakov01} using a compu\-tatio\-nal multi\--com\-part\-ment model for a representative synaptic geometry based on EM images.
To this end, we set $\kappa_{\{1,2\},y}=8.3\times 10^{-5}$~$\si{\micro\meter\per\micro\second}$ in agreement with \cite{rusakov01} and $k_{\{1,2\},z}=0$~$\si{\micro\meter\squared\per\micro\second}$, such that the particle flux at the boundaries $z=0$ and $z=l_z$ in our model represents the efflux of \acp{NT} from the synaptic space and the boundaries $y=0$ and $y=l_y$ represent the glial cell membrane.
We fix the synaptic cleft width, setting $l_x=0.02$~$\si{\micro\meter}$, and adjust $l_z$ such that the area of the glial cell membrane in our cuboid model, given by $2 l_x l_z$, corresponds to the inward facing surface area of the glial cell from the model synapse from \cite[Fig.~2(A2)]{rusakov01}, given appoximately as $2 \pi 0.225^2$~$\si{\micro\meter\squared}$, and set $l_y = l_z$, because half of this synapse is covered by glia.
To reproduce the configuration for the synapse ensheathed by ~95\% with glia from \cite[Fig.~2(A3)]{rusakov01}, we adjust $l_y$ and $l_z$ such that the volume of the cuboid does not change (as compared to the previous configuration), but the glial surface makes up 95\% of the total surface area in $y$ and $z$.
According to \cite{rusakov01}, we set $\kappa_r = 0.0$, $\kappa_d = 0.0$, and $\kappa_a$ close to zero.
Finally, we evaluate the particle concentration given in Proposition~\ref{prop:C} at the center of the synapse for these two scenarios (rescaling it to molar concentration \si{\milli M}).
The results are presented in Fig.~\ref{fig:rusakov_clements} together with data points obtained from \cite{rusakov01}.
It can be seen that our model predicts a biphasic concentration profile with fast decay in the first \si{\milli\second} and slower decay for $t > 1\si{\milli\second}$ in qualitative agreement with the concentration profile predicted by the model in \cite{rusakov01}.
For the synapse ensheathed by 50\% (brown curve in Fig.~\ref{fig:rusakov_clements}), our model slightly overestimates the intra-synaptic glutamate concentration as compared to \cite{rusakov01}.
For the synapse ensheathed by 95\% (blue curve in Fig.~\ref{fig:rusakov_clements}), our results agree even quantitatively very well with the results reported in \cite{rusakov01}.

Next, we compare Proposition~\ref{prop:C} with experimental results reported in \cite{clements96}.
In \cite{clements96}, the synaptic \ac{NT} concentration was studied experimentally by applying a rapidly dissociating receptor antagonist.
In lack of structural information \ac{wrt}~the synapse investigated in \cite{clements96}, we infer the model parameters from the biphasic concentration profile reported in \cite[Fig.~5]{clements96}.
From the diffusion dominated concentration decay in the first \si{\milli\second}, we extrapolate the effective synaptic volume available for free diffusion and set $l_y=l_z=1.1$~$\si{\micro\meter}$.
From the asymptotic half-life of \acp{NT} reported in \cite[Fig.~5]{clements96}, we infer $\kappa_{\{1,2\},\{y,z\}}=6\times 10^{-4}$~$\si{\micro\meter\per\micro\second}$.
Comparing the \ac{NT} concentration predicted by Proposition~\ref{prop:C} and shown as green curve in Fig.~\ref{fig:rusakov_clements} with the data points obtained from \cite[Fig.~5]{clements96}, we observe excellent quantitative agreement.

As we model the postsynaptic signal processing only up to the binding (and unbinding) of molecules to receptors and do not attempt to encompass further processing steps that would ultimately lead to the generation of an experimentally detectable electrical downstream signal, the comparison with experimental data {\em in terms of the \ac{CIR}} cannot be done strictly quantitatively.
However, we note that the \ac{CIR} predicted by our model is indeed plausible when compared to previous experimental and simulation studies.
In particular, the results presented in Section~\ref{sec:results:glial_cell_uptake}, Fig.~\ref{fig:glial_uptake}, suggest that the postsynaptic signal peaks within the first $0.5$~$\si{\milli\second}$ and vanishes after approximately $5$ to $6$~$\si{\milli\second}$.
These values are in good agreement with the rise and decay times of the postsynaptic signal found experimentally in \cite{savtchenko13} and by means of computer simulation in \cite[Fig.~3(C)]{nielsen04}.

\subsection{Presynaptic Reuptake and Postsynaptic Desorption for Reflective Case}\label{sec:results:reuptake}
\begin{figure*}[!tbp]
    \centering
    \begin{minipage}[t]{0.49\textwidth}
        \hspace{-0.15cm}	
        \includegraphics[width=\textwidth]{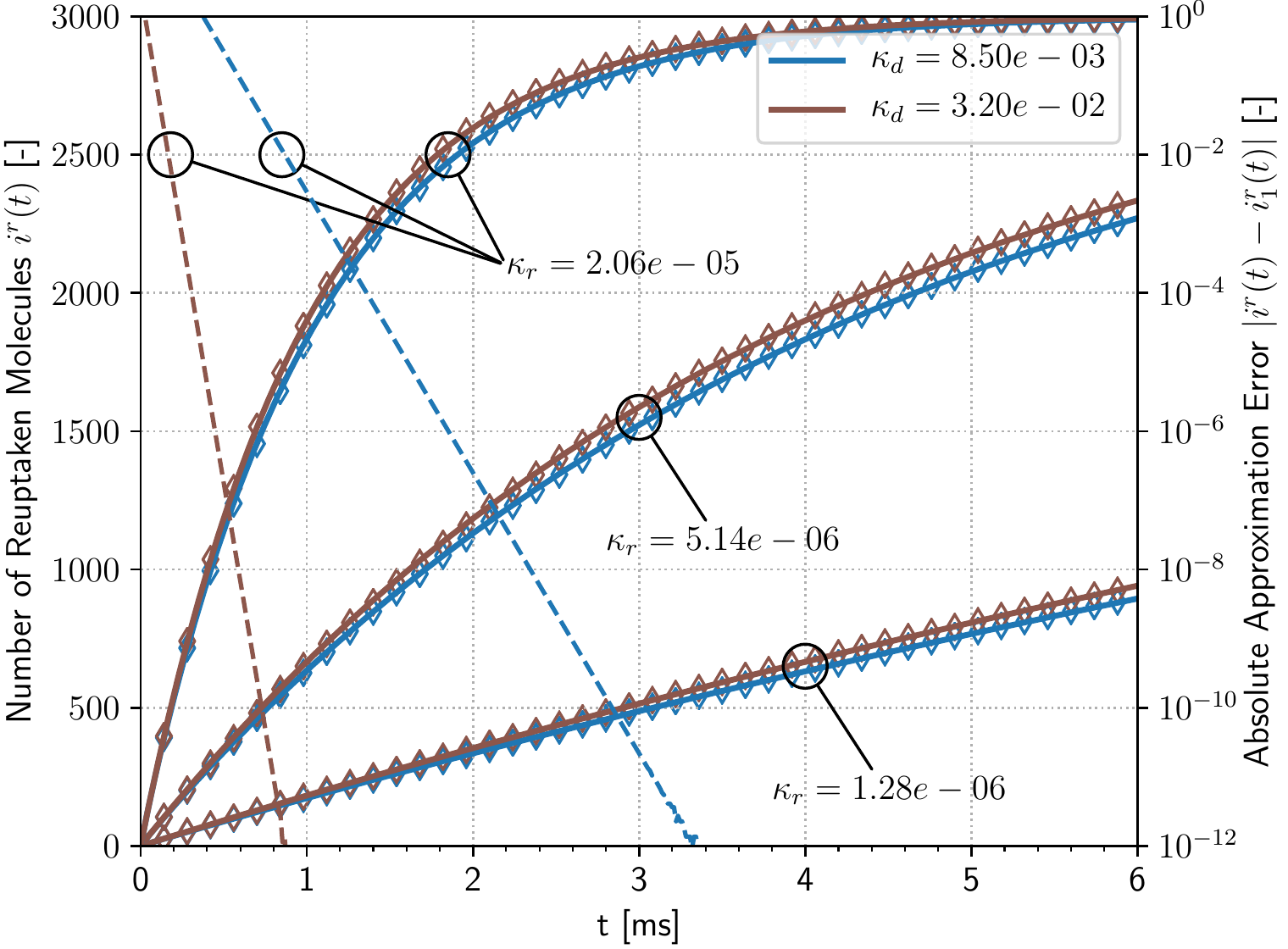}\vspace{-4mm}
        \caption{Left $y$-axis: Presynaptic reuptake in reflective case for different reuptake rates $\hlx$ in $\si{\micro\meter\per\micro\second}$ and different desorption rates $\hdx$ in $\si{\per\micro\second}$. The results from the particle-based simulations are depicted as diamond markers. Right $y$-axis: The absolute error of the first-term approximation \eqref{eq:ir_first_term} is shown as dashed lines.}
        \label{fig:reuptake_it}
    \end{minipage}
    \hfill
    \begin{minipage}[t]{0.49\textwidth}
        \hspace{-0.15cm}	
        \includegraphics[width=\textwidth]{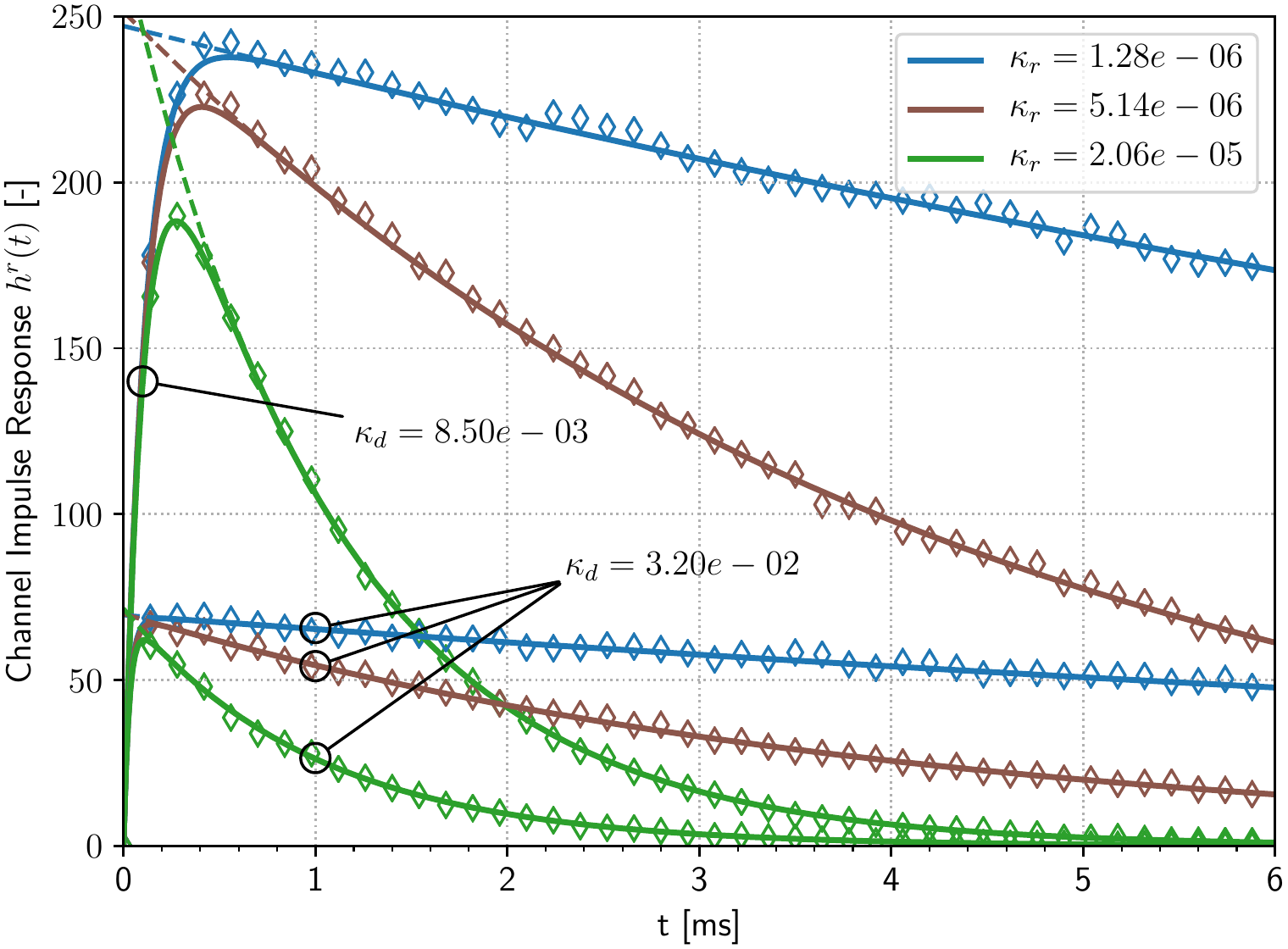}\vspace{-4mm}
        \caption{\ac{CIR} for different presynaptic reuptake rates $\hlx$ in $\si{\micro\meter\per\micro\second}$ and different desorption rates $\hdx$ in $\si{\per\micro\second}$ in the reflective case. The first-term approximation as defined in \eqref{eq:hr_first_term} is shown as dashed lines. The results from the particle-based simulations are shown as diamond markers.}
        \label{fig:reuptake_ht}
    \end{minipage}
    \vspace{-8mm}
\end{figure*}

In Figs.~\ref{fig:reuptake_it} and \ref{fig:reuptake_ht}, the impact of presynaptic reuptake and postsynaptic desorption on the \ac{CIR} and on the number of presynaptically reuptaken particles, respectively, in a synapse with reflective boundaries in $y$ and $z$ is shown.

From Fig.~\ref{fig:reuptake_it}, we observe that the time required by the presynaptic neuron to reuptake a given number of released particles decreases with increasing reuptake rate $\hlx$.
The desorption rate $\hdx$ on the other hand does not impact the presynaptic reuptake significantly.

Fig.~\ref{fig:reuptake_ht} shows how the presynaptic reuptake and the postsynaptic desorption shape the postsynaptic signal.
As the presynaptic reuptake $\hlx$ increases, the \ac{CIR} decays faster and its peak value decreases slightly.
This observation underlines the role of molecule reuptake for \ac{ISI} mitigation.
Increasing the desorption rate $\hdx$, on the other hand, leads to a significant decrease of the postsynaptic signal strength.

Hence, we conclude that presynaptic reuptake impacts both, the presynaptic and postsynaptic sides, while the effect of desorption is only local, at the postsynaptic side.

We further observe in Figs.~\ref{fig:reuptake_it} and \ref{fig:reuptake_ht} that the first-term approximation as proposed in \eqref{eq:ir_first_term} and \eqref{eq:def_h_first_term}, respectively, provides an excellent approximation to quickly evaluate the reuptake dynamics and the \ac{CIR} decay.
For the presynaptic reuptake, the approximation is accurate even for very small values of $t$.
In fact, we observed a significant decrease in the approximation quality only when pathological scenarios, such as a very small cleft width, were considered.

For the parameters considered here, $\hdx \frac{\hlx}{\hlx+\hrx} \ll \frac{\pi^2}{8}\frac{1}{t_{\lx}}$ and, according to Subsection \ref{sec:cir_decay_rate:reflective_case} and \eqref{eq:alpha_1_bounds}, the \ac{CIR} decay is limited by the reaction kinetics.
Note that, although in Fig.~\ref{fig:reuptake_ht} the signal seems to decay slower when $\hdx$ is increased, actually only the constant in front of the exponential in \eqref{eq:hr_first_term} decreases, but the decay rate of the exponential itself increases in accordance with the results from Section \ref{sec:cir_decay_rate}.

\subsection{Molecule Uptake at Glial Cells}\label{sec:results:glial_cell_uptake}
In this section, we explore how the morphology and the reaction kinetics of the glial cell impact the \ac{CIR} and the total number of uptaken particles.
This question is of practical importance because both parameters are variable in nature and have been shown to impact the synaptic channel clearance \cite{sweeney17}.

We consider the default set of parameters as listed in Table \ref{tab:sim_params} as baseline and then increase the distance to the glial cell, reflected in our model by $\ly$ and $\lz$, first by a factor of $1.5$ to $\ly=\lz=0.23$~$\si{\micro\meter}$ and then by a factor of $2$ to $\ly=\lz=0.3$~$\si{\micro\meter}$.
Similarly, we increase the glial cell uptake rate constant first by a factor of $1.5$ to $\hly=\hry=\hlz=\hrz=3.85 \times 10^{-5}$~$\si{\micro\meter\per\micro\second}$ and then by a factor of $2$ to $\hly=\hry=\hlz=\hrz=5.14 \times 10^{-5}$~$\si{\micro\meter\per\micro\second}$.

\begin{figure*}[!tbp]
    \centering
    \begin{minipage}[t]{0.49\textwidth}
        \hspace{-0.15cm}	
        \includegraphics[width=\textwidth]{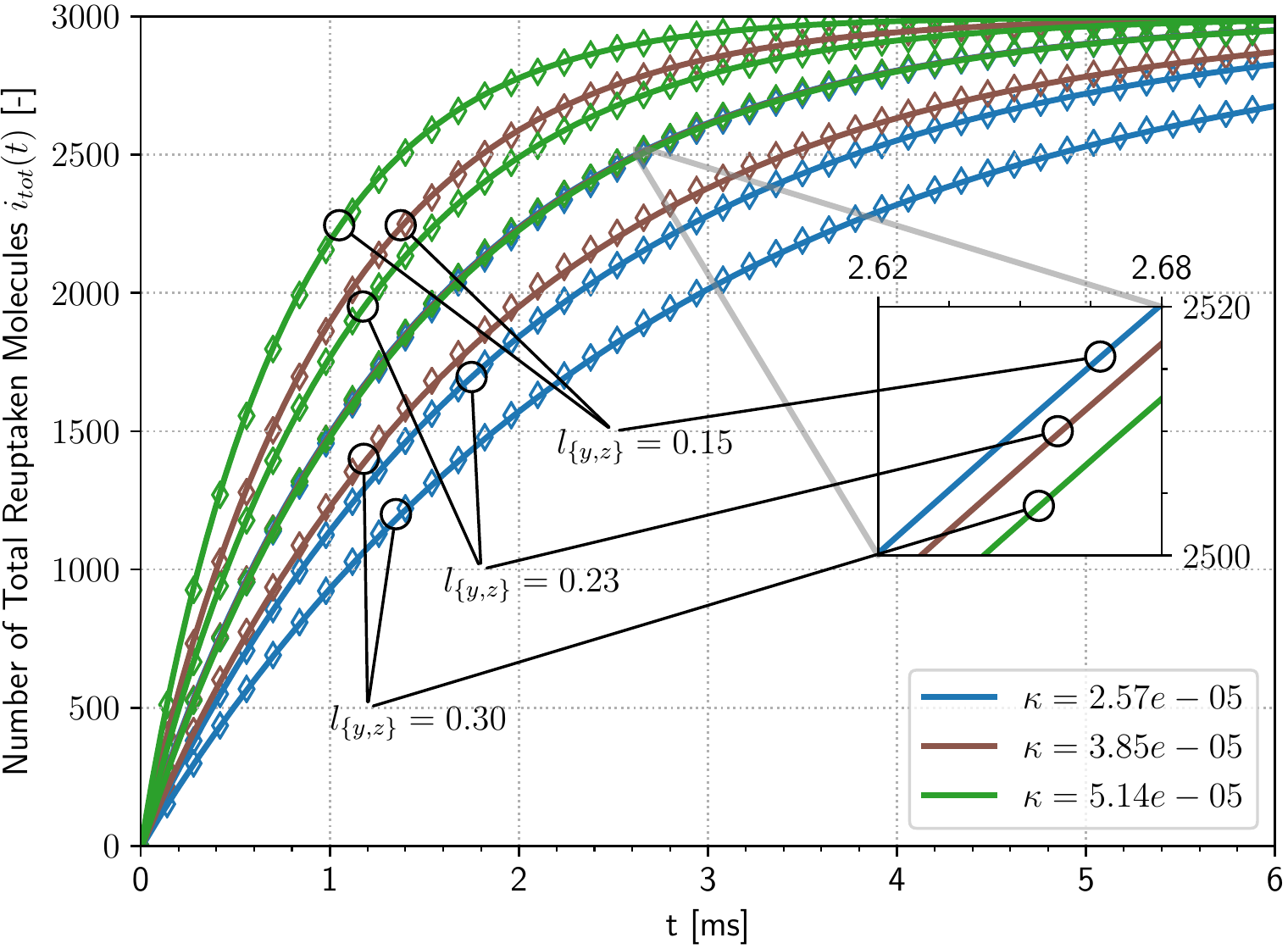}\vspace{-4mm}
        \caption{Total number of reuptaken particles for different distances to the glial cell $\ly$, $\lz$ in $\si{\micro\meter}$, and different glial cell uptake rates $\hly=\hry=\hlz=\hrz=\kappa$ in $\si{\micro\meter\per\micro\second}$. 
            The results from the particle-based simulations are depicted by the diamond markers.}
        \label{fig:glial_uptake_total_uptake}
    \end{minipage}
    \hfill
    \begin{minipage}[t]{0.49\textwidth}
        \hspace{-0.15cm}	
        \includegraphics[width=\textwidth]{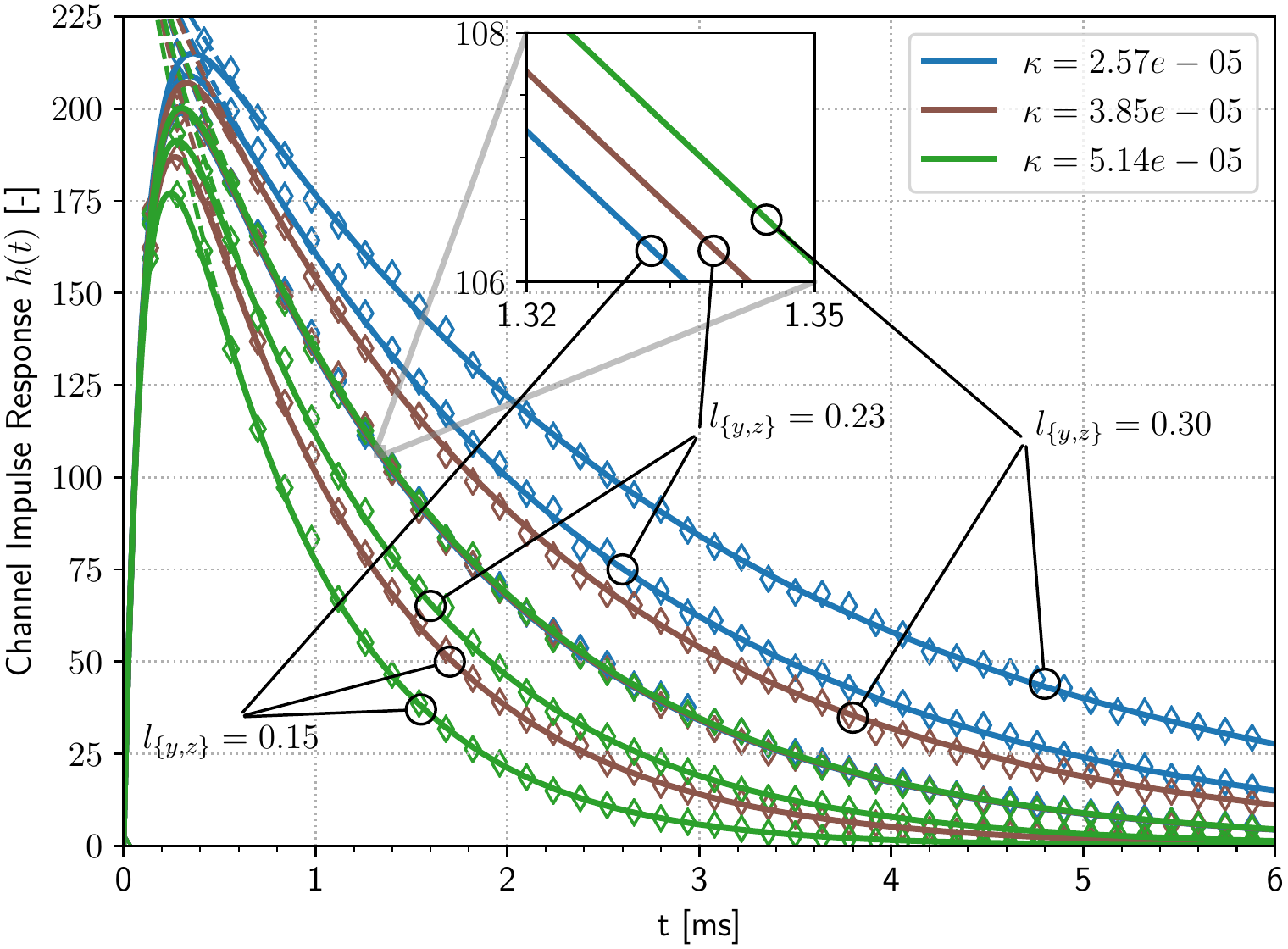}\vspace{-4mm}
        \caption{\ac{CIR} for different distances to the glial cell $\ly$, $\lz$ in $\si{\micro\meter}$, and different glial cell uptake rates $\hly=\hry=\hlz=\hrz=\kappa$ in $\si{\micro\meter\per\micro\second}$. 
            The results from the particle-based simulations are depicted by the diamond markers.
            The first-term approximation \eqref{eq:def_h_first_term} is shown as dashed lines.}
        \label{fig:glial_uptake}
    \end{minipage}
    \vspace{-8mm}
\end{figure*}

The results are shown in Figs.~\ref{fig:glial_uptake_total_uptake} and \ref{fig:glial_uptake}.
We observe that the effects of increasing the distance from the transmitter release site to the glial cell and increasing the uptake rate at the glial cell on the \ac{CIR} are opposite.
First, let us compare curves corresponding to low, medium, and large distance to the glial cell in Fig.~\ref{fig:glial_uptake}.
Clearly, the further the distance from release site to glial cell, the larger is the peak value of the \ac{CIR} and the slower it decays.
This is intuitive, because, as the distance to the glial cell membrane increases, more signaling molecules can reach the postsynaptic membrane before they are taken up by the glial cell.
On the other hand, comparing the results for low, medium, and high uptake rates in Fig.~\ref{fig:glial_uptake}, we observe that increasing the uptake rate clearly shortens the \ac{CIR} and, at the same time results in a smaller peak value.
Fig.~\ref{fig:glial_uptake_total_uptake} confirms that the different \ac{CIR} shapes actually have their correspondence in the shapes of the molecule uptake.
Here, {\em larger} distances lead to {\em slower} uptake, while {\em higher} uptake rates lead to {\em faster} uptake.

Now, we consider the effect of the simultaneous change of parameters.
First, we observe from Fig.~\ref{fig:glial_uptake} that the simultaneous increase of distance and uptake rate {\em by the same factor} does {\em not} impact the \ac{CIR} shape, i.e., the curves corresponding to $(l_{y,z},\kappa)=\{(0.15,2.57 \times 10^{-5}),(0.23,3.85 \times 10^{-5}),(0.3,5.14 \times 10^{-5})\}$, where $\kappa = \hly = \hry = \hlz = \hrz$, practically coincide.
The same applies to the number of uptaken particles as shown in Fig.~\ref{fig:glial_uptake_total_uptake}.
Thus, in the regime we are considering, the distance to the glial cell and the glial cell uptake rate can be traded-off one for the other.

On the other hand, if we compare the curves in Fig.~\ref{fig:glial_uptake} corresponding to $(l_{y,z},\kappa)=\{(0.23,2.57 \times 10^{-5}),(0.3,3.85 \times 10^{-5})\}$, we observe that increasing the glial cell uptake rate can also {\em overcompensate} for an increased distance to the glial cell.
Here, the distance is increased by a factor of $1.3$, while the uptake rate is increased by a factor of $1.5$ relative to the default values.
Most interestingly, we observe that the peak value of the \ac{CIR} remains unchanged while the \ac{CIR} decays faster.
This is because when increasing the distance, molecules will typically hit the glial cell membrane {\em later}, but will then be taken up {\em faster}.
This finding suggests that strong molecule uptake by the glial cell provides a potential mechanism for selectively suppressing the undesired tail of the postsynaptic signal making it a promising candidate for \ac{ISI} mitigation in synthetic systems, especially compared to less selective mechanisms like enzymatic degradation \cite{noel14}.

Fig.~\ref{fig:glial_uptake} shows also that the first-term approximation as defined in \eqref{eq:def_h_first_term} provides an accurate approximation to $h(t)$ shortly after the \ac{CIR} has peaked.
For $t > 0.5$~$\si{\milli\second}$ it is not distinguishable from $h(t)$.

A natural question arising from the preceding discussion is, to which extent increasing the glial cell uptake rate can speed up the \ac{CIR} decay.
But here, the bound for $\beta_1$ derived in Proposition \ref{prop:beta_1_bounds} comes in; increasing the glial cell uptake rate has only an effect on the \ac{CIR} decay as long as the diffusion-induced upper bound from \eqref{eq:def_bm_diff_lim_reg} is not reached.
Hence, if the glial cells are far apart, increasing the uptake rate does not significantly increase the \ac{CIR} decay.

\subsection{Vesicle Release Location and Spillover}\label{sec:results:release_location}
As explained in \cite[Section 3.2.6]{danbolt01}, the ability of glial cells to mitigate intersynaptic cross-talk, i.e., co-channel interference, which is due to the escape of signaling molecules from the synapse and their diffusive migration to neighboring synapses, highly depends on the position of the glial cell transporters relative to the molecule release site.
Here, we investigate this phenomenon by modeling spillover to neighboring synapses as particle flux over the boundaries $z=0$ and $z=\lz$.
In particular, we make the assumption that particles do not come back to the synapse once they have escaped from it and model spillover in this way as irreversible perfect adsorption.
Hence, we set $\klz=\krz=0$ to make the boundaries in $z$ perfectly adsorbing.

In reality, the distance from a particular synapse to neighboring synapses depends highly on the type of synapse under consideration and even within the same type is subject to variations by more than a factor of $100$ \cite{ventura99}.
Thus, we choose the location of our ``virtual'' boundaries, $\lz$, such that the contributions of glial cell molecule uptake in $y$ and spillover in $z$ to the overall molecule concentration decay, $\lambda^{1,1}_1$ as defined in \eqref{eq:def_lnmp}, are approximately equal and obtain $\lz = \SI{2}{\micro\meter}$.
According to \cite[Fig.~5]{ventura99} this is indeed a biologically plausible value.
Furthermore, to keep glial cell molecule uptake consistent with the default parameter values, the glial cell uptake rate constant is doubled to $\hly = \hry = 5.2 \times 10^{-5}$~$\si{\micro\meter\per\micro\second}$.

Since we are interested in the channel clearance due to glial cell uptake and there is a glial cell membrane only in $y$, we consider the molecule uptake in $y$, obtained from \eqref{eq:def_ig} as
\begin{align}
i_{y}(t) &= \sum_{n=1}^{\infty}\sum_{m=1}^{\infty}\sum_{p=1}^{\infty} \mathcal{A}_n^{m,p} \mathcal{B}_m \mathcal{C}_p C_0(x_0,y_0,z_0) \left[D \sin(\anmp \lx) + \frac{\hlx}{\anmp} \left(1 - \cos(\anmp \lx)\right)\right]\nonumber\\
&\times \left[\hly C_{m,y}(0) + \hry C_{m,y}(\ly)\right] \frac{\left(1 - \cos(\gp \lz)\right)}{\gp} \frac{\left(1-\exp(-\lnmp t)\right)}{\lnmp}.
\end{align}
as figure of interest.

\begin{figure*}[!tbp]
    \centering
    \begin{minipage}[t]{0.49\textwidth}
        \hspace{-0.15cm}	
        \includegraphics[width=\textwidth]{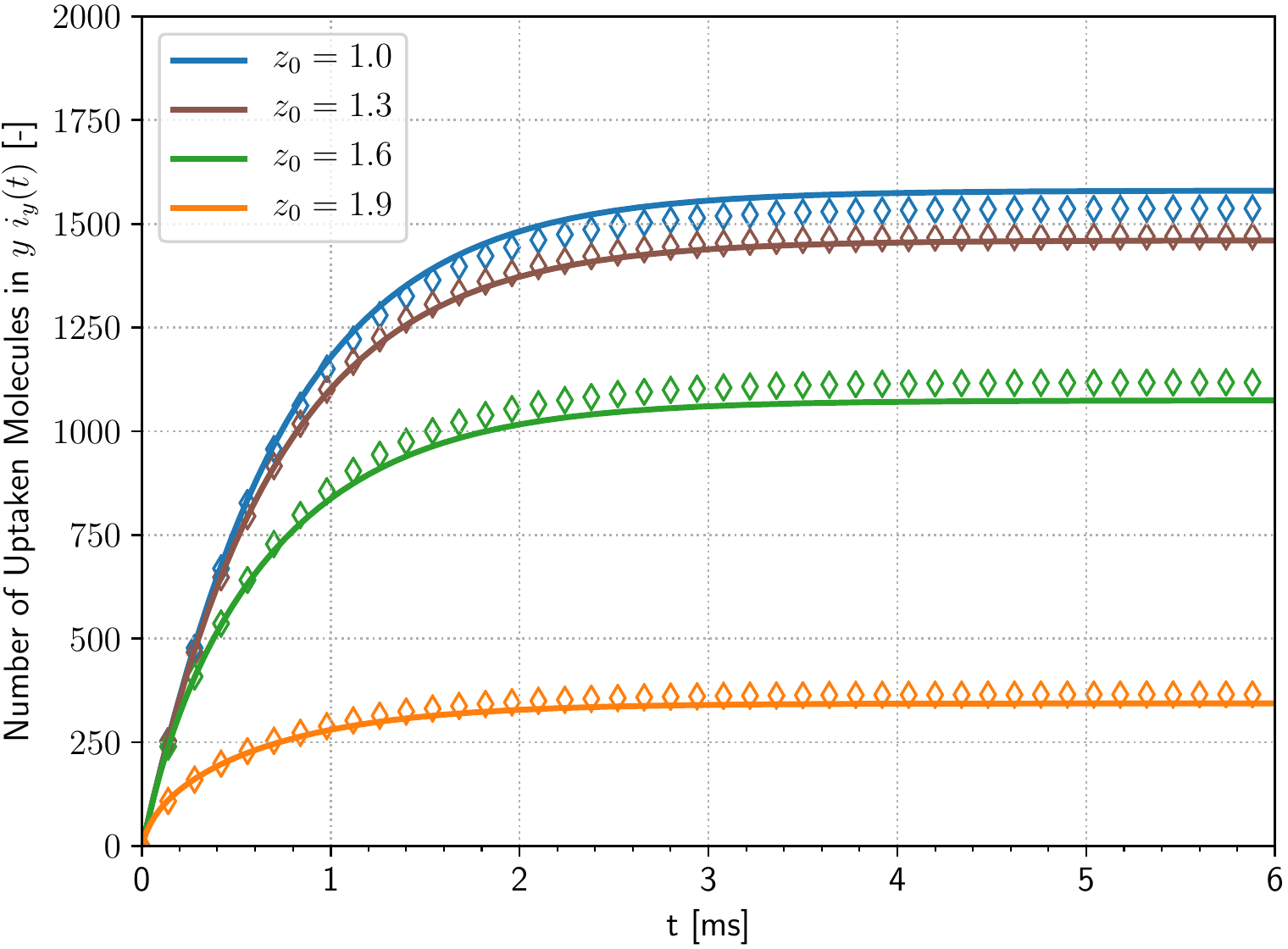}\vspace{-4mm}
        \caption{Number of uptaken particles at the glial cell in the presence of spillover for different release positions $z_0$ in $\si{\micro\meter}$. The simulation data is depicted by diamond markers.}
        \label{fig:release_position}
    \end{minipage}
    \hfill
    \begin{minipage}[t]{0.49\textwidth}
        \hspace{-0.15cm}	
        \includegraphics[width=\textwidth]{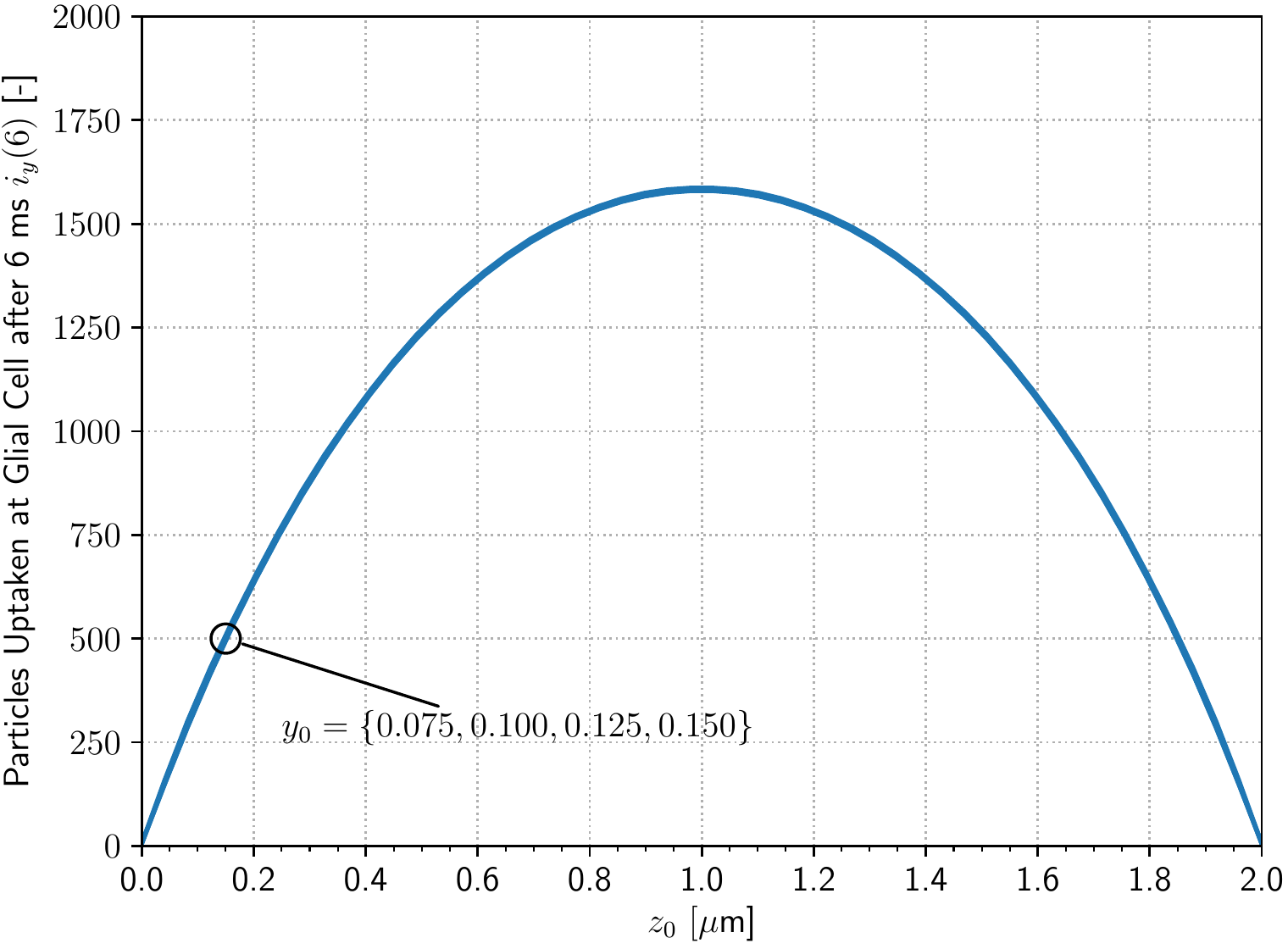}\vspace{-4mm}
        \caption{Number of uptaken particles at the glial cell after $6$~$\si{\milli\second}$ in the presence of spillover for different release positions $y_0$, $z_0$ in $\si{\micro\meter}$.}
        \label{fig:release_position_max_over_tx_z}
    \end{minipage}
    \vspace{-8mm}
\end{figure*}

The results are depicted in Figs.~\ref{fig:release_position} and \ref{fig:release_position_max_over_tx_z}.

In Fig.~\ref{fig:release_position}, we consider different release positions by varying the release position in $z$, $z_0$ in $\si{\micro\meter}$.
Note that, in contrast to the results presented in Subsections \ref{sec:results:reuptake} and \ref{sec:results:glial_cell_uptake}, the match between simulation data and analytical results is not perfect.
This is a consequence of the simulator design which does not allow for {\em exact} simulation of perfect adsorption, see \cite{clifford86} for a detailed analysis.
However, the agreement between simulation data and analytical results is still very good given that the evaluated quantities are cumulative, i.e., errors accumulate.

In Fig.~\ref{fig:release_position_max_over_tx_z}, the number of particles taken up at the glial cell after $6$ $\si{\milli\second}$ is shown depending on the release position in $z$, $z_0$, for different values of $y_0$.
First, we observe from Fig.~\ref{fig:release_position_max_over_tx_z} that the number of particles taken up at the glial cell is independent of $y_0$.
This is due to the fact that the reaction kinetics at the glial cell membrane are very slow compared to the diffusive motion of the particles; $\beta_1$ is in fact very close to the reaction-induced lower bound from \eqref{eq:def_bm_reac_lim_reg_a}, such that the spatial concentration profile in $y$ approaches the uniform distribution before the uptake exhibits a significant impact on the molecule concentration.
Therefore, it is not relevant for the molecule uptake at the glial cell whether particles are released close to the glial cell membrane or not.

In $z$ direction, on the other hand, due to the instantaneously absorbing boundary, particle diffusion rather than reaction kinetics limits the number of escaping particles and thus also the number of uptaken particles.
Indeed, $\gamma_1$ reaches the upper bound in \eqref{eq:def_bm_diff_lim_reg} and hence the system is diffusion-limited in $z$.
In accordance with that, we observe from Figs.~\ref{fig:release_position} and \ref{fig:release_position_max_over_tx_z} that the number of particles taken up at the glial cell heavily depends on the release position relative to the (virtual) system boundaries in $z$.
In particular, we observe from Fig.~\ref{fig:release_position_max_over_tx_z} that, as we increase or decrease $z_0$ towards $\lz$ or $0$, respectively, i.e., releasing the molecules closer to one of the absorbing non-glial boundaries, $i_{y}(t)$ decays the faster the closer the release site is to one of the boundaries, suggesting a nonlinear dependence of $i_{y}(t)$ on $z_0$.

\section{Conclusion}
\label{sec:conclusion}
In this paper, we have proposed an analytical time domain expression for the \ac{CIR} and the number of uptaken particles in the tripartite synapse.
The proposed model incorporates reversible binding to individual postsynaptic receptors, molecule uptake at the presynaptic cell and at glial cells, and spillover to the extrasynaptic space.
We have demonstrated the usefulness of our model by deriving fundamental bounds on the decay rate of the \ac{CIR} in dependence on channel geometry and chemical reaction kinetics.
Furthermore, it was shown that the model allows novel quantitative insights regarding the mitigation of \ac{ISI} and the effectiveness of \ac{EH} in synaptic communication assisted by a glial cell.
Although we have presented some exemplary evaluations, we believe that, by informed use of this model, new input to many open questions regarding signal transmission in the tripartite synapse can be obtained and further contributions towards the design of synthetic neuronal communication systems and \ac{DMC} systems in general are facilitated.
In particular, we expect novel insights from the results derived in this paper \ac{wrt}~\ac{ISI} mitigation and \ac{EH} in the tripartite synapse for a better understanding of the synaptic channel capacity.

Finally, we mention some limitations of the presented approach, which may motivate future research.
First, the assumption that synaptic signal transmission can be modeled as a linear time-invariant system breaks down as soon as any non-linear or time-dependent effects are considered.
Thus, the model is not capable of describing short and long term synaptic plasticity or time-dependent release due to activation of presynaptic auto-receptors \cite{zucker14}.
Second, in the quest for a realistic transmitter model, the study of the impact of (dysfunctional) \ac{NT} uptake on presynaptic release would nicely complement the results presented in this study \cite{hasegawa06}.
Third, the model that we used for postsynaptic receptors, is quite simplistic.
A more elaborate model would consider receptors with multiple binding sites, multiple states, and state-dependent kinetics \cite{jonas93}.
Finally, it would be interesting to investigate the impact of receptor occupancy, i.e.,~the fact that a molecule cannot bind to a receptor binding side which is already occupied by some other molecule \cite{mcallister00}.

\ifx \arXiv \undefined
\else
\appendix
\section{}
\ifx \arXiv \undefined
Due to space constraints, we provide only a sketch of the proof.
The full proof can be found in the version of this paper published at arXiv.
\fi
\subsection{Derivation of the CIR in the Laplace Domain}
\ifx \arXiv \undefined
Applying the eigenfunction expansion method explained in \cite{cain05} to the boundary-value problem \labelcref{eq:diff_eq,eq:pab_lb,eq:pab_rb,eq:rb_l,eq:rb_r,eq:pab_ic} yields the following equivalent formulation
\else
Let us first consider the eigenfunctions $\{\phi_m,\psi_p\}$ of the Sturm-Liouville problems \cite{cain05}
\begin{align}
\frac{\partial^2{}}{\partial y^2}\phi &= \beta \phi, 0 \leq y \leq \ly,\label{eq:eig_y}\\
\frac{\partial^2{}}{\partial z^2}\psi &= \gamma \psi, 0 \leq z \leq \lz,\label{eq:eig_z}
\end{align}
subject to
\begin{align}
\kly \frac{\partial \phi}{\partial y} &= \hly \phi \quad \textrm{at } y=0 & \kry \frac{\partial \phi}{\partial y} &= -\hry \phi \quad \textrm{at } y=\ly,\label{eq:eig_y_boundary}\\
\klz \frac{\partial \psi}{\partial z} &= \hlz \psi \quad \textrm{at } z=0 & \krz \frac{\partial \psi}{\partial z} &= -\hrz \psi \quad \textrm{at } z=\lz,\label{eq:eig_z_boundary}
\end{align}
where $\phi=\phi(y)$ and $\psi=\psi(z)$ are not identical $0$.
From \cite[Thm.~3.3]{cain05}, we know that $\{\phi_m\}$ and $\{\psi_p\}$ are countable sets and form an orthogonal basis of $L_2([0;l_y])$ and $L_2([0;l_z])$, respectively, where $L_2([a;b])$ denotes the set of square-integrable functions on the interval $[a;b]$, i.e., functions $f$, such that
\begin{align}
\|f\|^2_{[a;b]} := \int_{a}^{b}|f(x)|^2\mathrm{d}x < \infty.
\end{align}
Therefore, in particular, we can expand $C$ in these bases.
Assuming without loss of generality that $\{\phi_m\}$ and $\{\psi_p\}$ are normalized, i.e.,~$\|\phi_m\|^2_y=1$ and $\|\psi_p\|^2_z=1$, this is done in the usual way by orthogonal projection \cite{cain05}, which is defined as
\begin{align}
\mathrm{P}_y g &= \sum_{m=1}^{\infty} b_m \phi_m,&
\mathrm{P}_z h &= \sum_{p=1}^{\infty} c_p \psi_p,\label{eq:proj}
\end{align}
where $g \in L_2([0;l_y])$, $h \in L_2([0;l_z])$, and the coefficients $b_m,c_p$ are obtained as
\begin{align}
b_m &= \int_{0}^{\ly} g \phi_m \dy, &
c_p &= \int_{0}^{\lz} h \psi_p \dz.\label{eq:proj_coeff}
\end{align}
Thus, if we apply \eqref{eq:proj} to \eqref{eq:diff_eq} and exploit \eqref{eq:eig_y}, \eqref{eq:eig_z}, we obtain
\fi
\begin{equation}
\sum_{m=1}^{\infty}\sum_{p=1}^{\infty} \left[\frac{\partial^2{}C_{m,p}}{\partial x^2} - \beta^2_m C_{m,p} - \gamma^2_p C_{m,p} - \frac{1}{D} \frac{\partial{}C_{m,p}}{\partial t}\right] \phi_m \psi_p = 0,\label{eq:diff_proj}
\end{equation}
where the coefficients $C_{m,p}=C_{m,p}(x,t)$ depend on $x$ and $t$\ifx \arXiv \undefined ~and $\{\btm,\phi_m\}$ and $\{\gp,\psi_p\}$ are pairs of eigenvalues and eigenfunctions of the Sturm-Liouville problems corresponding to \labelcref{eq:diff_eq,eq:rb_l,eq:rb_r} \cite{cain05}\fi.
\ifx \arXiv \undefined
\else
Now, because the $\{\phi_m\}$ are linearly independent and also the $\{\psi_p\}$ are linearly independent, \eqref{eq:diff_proj} implies that
\begin{equation}
\frac{\partial^2{}C_{m,p}}{\partial x^2} - \beta^2_m C_{m,p} - \gamma^2_p C_{m,p} - \frac{1}{D} \frac{\partial{}C_{m,p}}{\partial t} = 0 \quad \forall m,p.\label{eq:diff_proj_mp}
\end{equation}
\fi
Furthermore, $C_{m,p}$ is to fulfill boundary conditions \eqref{eq:pab_lb}, \eqref{eq:pab_rb} and initial condition
\begin{equation}
C_{m,p,0}=C_{m,p}(x,0)= \int_{0}^{\lz}\int_{0}^{\ly} C_0 \phi_m \psi_p \dy\dz = \delta(x-x_0) \phi_m(y_0) \psi_p(z_0).
\end{equation}
\ifx \arXiv \undefined
Since the $\{\phi_m\}$ are linearly independent and also the $\{\psi_p\}$ are linearly independent, the summands in \eqref{eq:diff_proj} must be zero for all $m,p \in \mathbb{N}$ and we can employ the Laplace transform of $C_{m,p}(x,t)$ with respect to $t$, defined as 
\begin{equation}
\bar{C}_{m,p}(x,s) = \mathcal{L}\{C_{m,p}(x,t)\} = \int_{0}^{\infty}C_{m,p}(x,\tau)\exp(-s\tau) \mathrm{d}\tau,
\end{equation}
and use the strategy from \cite[Appendix A]{lotter20a} to find
\begin{equation}
\bar{C}_{m,p} = \frac{\phi_m(y_0) \psi_p(z_0)}{2 D q'}e^{-q'|x-x_0|} + A \sinh(q'x) + B \cosh(q'x),\label{eq:sol_lapl}
\end{equation}
where
\else
Now, we apply the Laplace transform with respect to $t$, denoted by $\mathcal{L}$, to \eqref{eq:diff_proj_mp} and obtain the inhomogeneous ordinary differential equation
\begin{equation}
\frac{\partial^2{}\bar{C}_{m,p}}{\partial x^2} = \left(\frac{s}{D} + \beta^2_m + \gamma^2_p\right) \bar{C}_{m,p}  - \frac{1}{D} C_{m,p,0},\label{eq:diff_proj_lapl}
\end{equation}
where $s$ denotes the Laplace variable, $\bar{C}_{m,p}(x,s)=\mathcal{L}\{C_{m,p}(x,t)\}=\int_{0}^{\infty}C_{m,p}(x,\tau)\exp(-s\tau) \mathrm{d}\tau$, and \eqref{eq:diff_proj_lapl} is subject to
\begin{align}
D\frac{\partial \bar{C}_{m,p}}{\partial x} &= \kappa_r \bar{C}_{m,p},\quad & &\text{at } x = 0, \textrm{ and}\label{eq:pab_lb_lapl} \\
D \frac{\partial \bar{C}_{m,p}}{\partial x} &= -\kappa_a \bar{C}_{m,p} - \kappa_d D \frac{1}{s} \frac{\partial \bar{C}_{m,p}}{\partial x},\quad & &\text{at } x = \lx. \label{eq:pab_rb_lapl}
\end{align}
Here, we have exploited the properties of the Laplace transform that $\mathcal{L}\{\frac{\mathrm{d}f}{\mathrm{d}t}\} = s \mathcal{L}\{f\} - f(0)$ and $\mathcal{L}\{\int_{0}^{t}f\mathrm{d}\tau\} = \frac{1}{s}\mathcal{L}\{f\}$.
To solve \labelcref{eq:diff_proj_lapl,eq:pab_lb_lapl,eq:pab_rb_lapl}, we follow a similar approach as \cite[Ch.~14]{carslaw86} and decompose $C_{m,p}$ as
\begin{equation}
C_{m,p} = U_{m,p} + W_{m,p},
\end{equation}
such that both, $U_{m,p}$ and $W_{m,p}$, fulfill \eqref{eq:diff_proj_mp}, $U_{m,p}=C_{m,p,0}$ at $t=0$, $W_{m,p}=0$ at $t=0$, and the sum of $U_{m,p}$ and $W_{m,p}$ fulfills \eqref{eq:pab_lb} and \eqref{eq:pab_rb}.
Let us denote the Laplace transforms of $U_{m,p}$ and $W_{m,p}$ as $\bar{U}_{m,p}$ and $\bar{W}_{m,p}$, respectively, and define $q'=\sqrt{\frac{s}{D}+\btm^2+\gp^2}$.
Then, from \cite[Ch.~14.3]{carslaw86}, we find that
\begin{align}
\bar{U}_{m,p} &= \frac{\phi_m(y_0) \psi_p(z_0)}{2 D q'}e^{-q'|x-x_0|},\\
\bar{C}_{m,p} &= \bar{U}_{m,p} + \bar{W}_{m,p} = \frac{\phi_m(y_0) \psi_p(z_0)}{2 D q'}e^{-q'|x-x_0|} + A \sinh(q'x) + B \cosh(q'x),\label{eq:sol_lapl}
\end{align}
where $A$ and $B$ are constants with respect to $x$ and must be chosen such that \eqref{eq:pab_lb_lapl} and \eqref{eq:pab_rb_lapl} are fulfilled.
Substituting \eqref{eq:sol_lapl} in \eqref{eq:pab_lb_lapl} and \eqref{eq:pab_rb_lapl} yields
\fi
\begin{align}
A = &\frac{1}{q'}\left(K_{1,x} \left.\bar{U}_{m,p}\right|_{x=0} - \left.\frac{\partial \bar{U}_{m,p}}{\partial x}\right|_{x=0} + B K_{1,x}\right),\\
B = &\left[-\left.q'\frac{\partial \bar{U}_{m,p}}{\partial x}\right|_{x=\lx} + q'\left(\left.\frac{\partial \bar{U}_{m,p}}{\partial x}\right|_{x=0} - K_{1,x} \left.\bar{U}_{m,p}\right|_{x=0}\right)\cosh(q'\lx) -q'K_{2,x}\left. \bar{U}_{m,p}\right|_{x=\lx}\right.\nonumber\\
&{}-\left.\left(K_{1,x}K_{2,x}\left.\bar{U}_{m,p}\right|_{x=0} - K_{2,x}\left.\frac{\partial \bar{U}_{m,p}}{\partial x}\right|_{x=0}\right)\sinh(q'\lx)\right]/\nonumber\\
&\left[q'(K_{1,x}+K_{2,x})\cosh(q'\lx) + \left((q')^2+K_{1,x}K_{2,x}\right)\sinh(q'\lx)\right],
\end{align}
and we have set $q'=\sqrt{\frac{s}{D}+\btm^2+\gp^2}$, $K_{1,x}=\hlx/\klx$ and $K_{2,x}=\hrx/\left[\krx(1+\frac{\hdx}{s})\right]$.

\subsection{Time Domain Solution}
\ifx \arXiv \undefined
To transform \eqref{eq:sol_lapl} back to the time domain, we use the residue theorem as outlined in \cite[Appendix B]{lotter20a}.
For the non-zero residues $\sigma^{m,p}_n$ of $\bar{C}_{m,p}$ we obtain
\begin{align}
\textrm{Res}\left(e^{st}\bar{C}_{m,p},\sigma^{m,p}_n\right) &= \phi_m(y_0) \psi_p(z_0)(\hlx \sin(\anmp x_0) + \klx \anmp \cos(\anmp x_0))\\
&\times 2[(\anmp)^2 \krx^2 (\lnmp - \hdx)^2 + (\lnmp)^2 \hrx^2] / \nonumber \\
&\{(\lnmp - \hdx)[D^3 (\anmp)^2(\hlx (\lnmp - \hdx) + \hrx \lnmp) + \hlx^2 \hrx \krx \lnmp] + \nonumber \\
&+ 2 \hlx^2 \hrx \hdx D^2 (\anmp)^2 + \hlx \hrx^2 \klx (\lnmp)^2 + 2 \hrx \hdx D^4 (\anmp)^4 + \nonumber \\
&+l_x [\hlx^2 \krx^2 (\anmp)^2 (\lnmp - \hdx)^2 + \hrx^2 \klx^2 (\anmp)^2 (\lnmp)^2 + \nonumber \\
&+ D^4 (\anmp)^4 (\lnmp - \hdx)^2 + \hlx^2 \hrx^2 (\lnmp)^2]\}\\
&\times (\hlx \sin(\anmp x) + \klx \anmp \cos(\anmp x))\exp(-\lnmp t),
\end{align}
where $\sigma^{m,p}_n = -((\anmp)^2+\btm^2+\gp^2)D$ and $\anmp$, $\btm$, $\gp$, and $\lnmp$ are defined in \eqref{eq:def_an}, \eqref{eq:def_btm}, \eqref{eq:def_gp}, and \eqref{eq:def_lnmp}, respectively.
Furthermore, if $K_{1,x} = 0$ and $\btm^2 + \gp^2 = 0$, $\bar{C}_{m,p}$ has a double pole at $q'=0$ and the corresponding residue reads $\hdx(\phi_m(y_0) \psi_p(z_0))/(\hrx + \lx \hdx)$.
\else
Now, to transform \eqref{eq:sol_lapl} back to time domain, we need to compute the inverse Laplace transform,
\begin{equation}
C_{m,p}(x,t) = \frac{1}{2 \pi j} \int_{\nu -j \infty}^{\nu + j \infty} e^{st}\bar{C}_{m,p}(x,s) \textrm{d}s, \label{eq:inv_lapl_trans}
\end{equation}
where $j$ denotes the imaginary unit and $\nu \in \mathbb{R}^+$ needs to be chosen large enough such that all singularities of $\bar{C}_{m,p}$ are left of the line along which the integral is computed.

Similarly to \cite{carslaw86}, instead of integrating along the infinite line, we complete the integration path with a semicircle to a closed contour which contains the origin and all singularities of $\bar{C}_{m,p}$.
Then, we use the residue theorem to replace this contour integral with a sum over the residues $\{\sigma\}$ of $\bar{C}_{m,p}$,
\begin{equation}
C_{m,p}(x,t) = \sum_{\{\sigma\}} \mathrm{Res}(e^{st}\bar{C}_{m,p},\sigma). \label{eq:residue_sum}
\end{equation}

Now, \eqref{eq:sol_lapl} has poles at $q'=0$ and infinitely many non-zero simple poles defined as the solutions of
\begin{equation}
(K_{1,x}+K_{2,x})\cosh(q'\lx) + \left((q')^2+\frac{K_{1,x}K_{2,x}}{q'}\right)\sinh(q'\lx) = 0.\label{eq:simple_nonzero_pole}
\end{equation}
Note that, at $q'=0$, \eqref{eq:simple_nonzero_pole} does not have a pole but a removable singularity as $\lim\limits_{q' \to 0} \sinh(q'\lx)/q' = 1$.
If $K_{1,x}=0$ and $\btm=\gp=0$, \eqref{eq:simple_nonzero_pole} has a root at $q'=0$ and thus contributes a pole at $q'=0$ to \eqref{eq:sol_lapl}.

To compute \eqref{eq:residue_sum}, we start with the non-zero roots of \eqref{eq:simple_nonzero_pole}.

The residues of $e^{st}\bar{C}_{m,p}$ at simple poles $\sigma$ can be computed with l'H\^opital's rule as
\begin{equation}
\textrm{Res}\left(e^{st}\bar{C}_{m,p},\sigma\right) = \left.\frac{e^{st}\bar{C}^N_{m,p}}{\frac{\partial \bar{C}^D_{m,p}}{\partial s}}\right|_{s=\sigma}, \label{eq:res_lhopital}
\end{equation}
where $\bar{C}^N_{m,p}$ and $\bar{C}^D_{m,p}$ denote the numerator and denominator of $\bar{C}_{m,p}$, respectively, as defined by \eqref{eq:sol_lapl}.

Now, we set $q'=j \alpha$ and from \eqref{eq:simple_nonzero_pole} it follows that the non-zero poles of \eqref{eq:sol_lapl} correspond to the non-zero roots $\anmp$, $n \in \mathbb{N}$ of \eqref{eq:def_an} with either positive real part or positive imaginary part. 
Eq.~\eqref{eq:def_an} admits infinitely many pairs of real solutions and, under some conditions (if $\hdx > 0$ and $\btm^2+\gp^2$ is large enough), also one pair of complex conjugate pure imaginary roots.
Due to our substitution $q'=\sqrt{\frac{s}{D}+\btm^2+\gp^2}$, we need to consider only one root from each pair of roots and \ac{wlog} we choose the one with positive real or positive imaginary part, respectively.

Furthermore, we index the real roots ascendingly, starting with index $1$ if \eqref{eq:def_an} does {\em not} admit imaginary roots and starting with index $2$ otherwise.
In the latter case, we index the imaginary root of \eqref{eq:def_an} with index $1$.

Applying \eqref{eq:res_lhopital} to \eqref{eq:sol_lapl} we obtain
\begin{align}
\textrm{Res}\left(e^{st}\bar{C}_{m,p},\sigma^{m,p}_n\right) &= \phi_m(y_0) \psi_p(z_0)(\hlx \sin(\anmp x_0) + \klx \anmp \cos(\anmp x_0))\\
&\times 2[(\anmp)^2 \krx^2 (\lnmp - \hdx)^2 + (\lnmp)^2 \hrx^2] / \nonumber \\
&\{(\lnmp - \hdx)[D^3 (\anmp)^2(\hlx (\lnmp - \hdx) + \hrx \lnmp) + \hlx^2 \hrx \krx \lnmp] + \nonumber \\
&+ 2 \hlx^2 \hrx \hdx D^2 (\anmp)^2 + \hlx \hrx^2 \klx (\lnmp)^2 + 2 \hrx \hdx D^4 (\anmp)^4 + \nonumber \\
&+l_x [\hlx^2 \krx^2 (\anmp)^2 (\lnmp - \hdx)^2 + \hrx^2 \klx^2 (\anmp)^2 (\lnmp)^2 + \nonumber \\
&+ D^4 (\anmp)^4 (\lnmp - \hdx)^2 + \hlx^2 \hrx^2 (\lnmp)^2]\}\\
&\times (\hlx \sin(\anmp x) + \klx \anmp \cos(\anmp x))\exp(-\lnmp t),
\end{align}
where $\sigma^{m,p}_n = -((\anmp)^2+\btm^2+\gp^2)D$ and $\lnmp$ is defined in \eqref{eq:def_lnmp}.

Now, for the poles of \eqref{eq:sol_lapl} at $q'=0$, we employ a different strategy.
Namely, here we use the fact that the residue of a function at a pole $\sigma$ is given as the coefficient of the term of order $-1$ of its Laurent series expansion at $\sigma$.
Hence, if the Laurent series expansion of $e^{st}\bar{C}_{m,p}$ at $\sigma$ is given as
\begin{equation}
e^{st}\bar{C}_{m,p} = \sum_{k=-\infty}^{\infty}c_k (s - \sigma)^k,
\end{equation}
we have $\textrm{Res}(e^{st}\bar{C}_{m,p},\sigma) = c_{-1}$.

For $q'=0$, we have $s = -(\btm^2+\gp^2)D$ and thus we need the Laurent series expansion of $e^{st}\bar{C}_{m,p}$ at $\sigma = -(\btm^2+\gp^2)D$.
Before we perform the series expansion, we apply the change of variable $q'=\sqrt{\frac{s}{D}+\btm^2+\gp^2}$ and note that
\begin{equation}
(s - \sigma)^{-1} = (s + (\btm^2+\gp^2)D)^{-1} = D^{-1}(q')^{-2}.
\end{equation}
Thus, we can equivalently find the series expansion of $e^{st}\bar{C}_{m,p}$ at $q'=0$,
\begin{equation}
e^{st}\bar{C}_{m,p} = \sum_{k=-\infty}^{\infty} d_k (q')^{k},
\end{equation}
and have $\textrm{Res}(e^{st}\bar{C}_{m,p},\sigma) = c_{-1} = d_{-2} D$.

First, if $K_{1,x} \neq 0$, \eqref{eq:sol_lapl} has a simple pole at $q'=0$ and by expanding the expression in $q'=0$, it can easily be shown that in this case $\textrm{Res}(e^{st}\bar{C}_{m,p},\sigma) = 0$.
The same observation holds, if $K_{1,x} = 0$ but $\btm^2 + \gp^2 \neq 0$.

Now, if $K_{1,x} = 0$ and $\btm^2 + \gp^2 = 0$, there are non-zero contributions from only two terms in \eqref{eq:sol_lapl}, namely,
\begin{equation}
\frac{\left.\frac{\partial \bar{U}_{m,p}}{\partial x}\right|_{x=0} \cosh (q' \lx) \cosh(q' x)}{q'(\frac{K_{2,x}}{q'}\cosh (q' \lx)+\sinh (q' \lx))}\label{eq:residue_const_part_a}
\end{equation}
and
\begin{equation}
\frac{-\left.\frac{\partial \bar{U}_{m,p}}{\partial x}\right|_{x=\lx}\cosh(q' x)}{q'(\frac{K_{2,x}}{q'}\cosh (q' \lx)+\sinh (q' \lx))}.\label{eq:residue_const_part_b}
\end{equation}
For \eqref{eq:residue_const_part_a}, with $s = (q')^2 D$, we have
\begin{equation}
\frac{\left.\frac{\partial \bar{U}_{m,p}}{\partial x}\right|_{x=0} \cosh (q' \lx) \cosh(q' x)e^{(q')^2 D t}}{q'(\frac{K_{2,x}}{q'}\cosh (q' \lx)+\sinh (q' \lx))}=\frac{\phi_m(y_0) \psi_p(z_0)}{2 D (q')^2}\frac{\cosh (q' \lx) \cosh(q' x)e^{(q')^2 D t - q' x_0}}{(\frac{K_{2,x}}{(q')^2}\cosh (q' \lx)+\frac{\sinh (q' \lx)}{q'})}.\label{eq:residue_const_part_a_rw}
\end{equation}
Now, we denote the Taylor series expansion of the right-hand side of \eqref{eq:residue_const_part_a_rw} at $q'=0$ as
\begin{equation}
\frac{\cosh (q' \lx) \cosh(q' x)e^{q'(q' D t - x_0)}}{\frac{K_{2,x}}{(q')^2}\cosh (q' \lx)+\frac{\sinh (q' \lx)}{q'}} = d^{(1)}_0 + d^{(1)}_1 q' + d^{(1)}_2 (q')^2 + \ldots
\end{equation}
and find
\begin{align}
1 + O(q') &= \cosh (q' \lx) \cosh(q' x)e^{q'(q' D t - x_0)}\nonumber\\
&= (d^{(1)}_0 + O(q'))\left(\frac{K_{2,x}}{(q')^2}\cosh (q' \lx)+\frac{\sinh (q' \lx)}{q'}\right)\nonumber\\
&=(d^{(1)}_0 + O(q'))\left\lbrace\frac{\hrx}{\hdx}\frac{1}{\frac{(q')^2 D}{\hdx}+1}\left[1+O((q')^2)\right]+\left[\lx + O((q')^2)\right]\right\rbrace\nonumber\\
&=(d^{(1)}_0 + O(q'))\left\lbrace\frac{\hrx}{\hdx}\left[1 +O((q')^2)\right]\left[1+O((q')^2)\right]+\left[\lx + O((q')^2)\right]\right\rbrace,\label{eq:residue_const_part_a_exp}
\end{align}
where $O(\cdot)$ denotes big O notation and we have used the well-known series expansions of $e^x$, $\sinh(x)$, $\cosh (x)$, and $1/(x^2 + 1)$.
Collecting the constant terms in \eqref{eq:residue_const_part_a_exp}, we find
\begin{equation}
1 = d^{(1)}_0 \left(\frac{\hrx}{\hdx}+\lx\right),
\end{equation}
and therefore
\begin{equation}
d^{(1)}_0 = \frac{\hdx}{\hrx + \lx \hdx}.\label{eq:residue_constant_part_a_pre}
\end{equation}
Multiplying both sides of \eqref{eq:residue_const_part_a_exp} by $\phi_m(y_0) \psi_p(z_0)/(2 D (q')^2)$, we obtain the Laurent series expansion of the left-hand side of \eqref{eq:residue_const_part_a_rw} as
\begin{equation}
\frac{\left.\frac{\partial \bar{U}_{m,p}}{\partial x}\right|_{x=0} \cosh (q' \lx) \cosh(q' x)e^{(q')^2 D t}}{q'(\frac{K_{2,x}}{q'}\cosh (q' \lx)+\sinh (q' \lx))} = \frac{\hdx}{\hrx + \lx \hdx} \frac{\phi_m(y_0) \psi_p(z_0)}{2 D (q')^2} + O((q')^{-1})\label{eq:residue_const_part_a_fin}
\end{equation}
and in a similar fashion we obtain
\begin{equation}
\frac{-\left.\frac{\partial \bar{U}_{m,p}}{\partial x}\right|_{x=\lx}\cosh(q' x)e^{(q')^2 D t}}{q'(\frac{K_{2,x}}{q'}\cosh (q' \lx)+\sinh (q' \lx))} = \frac{\hdx}{\hrx + \lx \hdx} \frac{\phi_m(y_0) \psi_p(z_0)}{2 D (q')^2} + O((q')^{-1}).\label{eq:residue_const_part_b_fin}
\end{equation}
Finally, summing \eqref{eq:residue_const_part_a_fin} and \eqref{eq:residue_const_part_b_fin}, we conclude that $d_{-2} = \hdx(\phi_m(y_0) \psi_p(z_0))/\left[D(\hrx + \lx \hdx)\right]$ and thus $c_{-1} = \hdx(\phi_m(y_0) \psi_p(z_0))/(\hrx + \lx \hdx)$.
\fi

Putting together the pieces, we obtain
\begin{align}
C_{m,p}(x,t) &= \frac{\hdx}{\hrx + \lx \hdx} \phi_m(y_0) \psi_p(z_0) \ind{\kr}{0}\ind{\btm}{0}\ind{\gp}{0}\\
&+ \sum_{n=1}^{\infty} \mathcal{A}_n^{m,p} C_0(x_0,y_0,z_0) C_{n,x}^{m,p}(x) \exp(-\lnmp t),
\end{align}
where $\mathcal{A}_n^{m,p}$, $C_0(x_0,y_0,z_0)$, and $C_{n,x}^{m,p}(x)$ are defined in \eqref{eq:def_Anmp}, \eqref{eq:def_C0}, and \eqref{eq:def_Cmpnx}, respectively.

Now, to obtain the final solution $C(x,y,z,t)$ as a linear combination of orthonormal basis functions, we need to determine $\lbrace\phi_m\rbrace$ and $\lbrace\psi_p\rbrace$.
But these are well-known in the literature, see for example \cite{cain05,carslaw86}, and we repeat them here for completeness only:
\begin{align}
 \phi_m(y) &= \left\lbrace\frac{\left[2-\ind{\btm}{0}\right](\btm^2 \kry^2 + \hry^2)}{\left[l_y(\hly^2+\kly^2\btm^2)+\hly \kly\right](\hry^2+\kry^2\btm^2)+ \hry \kry (\hly^2+\kly^2\btm^2)}\right\rbrace^{\frac{1}{2}}\nonumber\\
 &\times \left[\hly \sin(\btm y) + \kly \btm \cos(\btm y)\right],\\
 \psi_p(z) &= \left\lbrace\frac{\left[2-\ind{\gp}{0}\right](\gp^2 \krz^2 + \hrz^2)}{\left[l_z(\hlz^2+\klz^2\gp^2)+\hlz \klz\right](\hrz^2+\krz^2\gp^2)+ \hrz \krz (\hlz^2+\klz^2\gp^2)}\right\rbrace^{\frac{1}{2}}\nonumber\\
 &\times \left[\hlz \sin(\gp z) + \klz \gp \cos(\gp z)\right],
\end{align}
where $\btm$ and $\gp$ are the positive, or, if $\hly=\hry=0$ $(\hlz=\hrz=0)$, the non-negative, roots of \eqref{eq:def_btm}, \eqref{eq:def_gp}.
All special cases, e.g.~perfectly absorbing boundaries $\kly=\kry=\klz=\krz=0$, follow from this definition.
In some special cases, \eqref{eq:def_btm}, \eqref{eq:def_gp} simplify considerably and an explicit expressions can be given for $\lbrace \btm \rbrace$ and $\lbrace \gp \rbrace$.
For a comprehensive listing of different combinations of boundary conditions and the associated eigenvalues, see for example \cite[Tab.~2.1]{ozisik13}.
\ifx \arXiv \undefined
If the boundary in $y$ ($z$) is reflective, i.e., $\hly=\hry=0$ $(\hlz=\hrz=0)$, the constant eigenfunction $1/\sqrt{\ly}$ ($1/\sqrt{\lz}$) is admitted and thus also the eigenvalue $\beta = 0$ ($\gamma = 0$) needs to be considered.
\else
If the boundary in $y$ ($z$) is reflective, i.e., $\hly=\hry=0$ $(\hlz=\hrz=0)$, the constant eigenfunction $1/\sqrt{\ly}$ ($1/\sqrt{\lz}$) is admitted by \eqref{eq:eig_y}, \eqref{eq:eig_y_boundary} (\eqref{eq:eig_z}, \eqref{eq:eig_z_boundary}), and thus also the eigenvalues $\beta = 0$ ($\gamma = 0$) need to be considered.
\fi

This completes the proof.

\fi

\bibliographystyle{IEEEtran}
\bibliography{IEEEabrv,H:/libraries/mc}
\end{document}